\documentclass[authoryear,final,3p,12pt]{elsarticle}

\usepackage{bbm}
\usepackage{url}
\usepackage{tabularx}
\usepackage{colortbl,booktabs}
\usepackage{multirow}
\usepackage{threeparttable}
\usepackage{makecell}
\usepackage{algorithm}
\usepackage{verbatim}
\usepackage{algorithmic}
\usepackage{xcolor}
\usepackage{diagbox}
\usepackage{enumerate}
\usepackage{threeparttable}
\usepackage{graphicx}
\usepackage{hyperref}
\usepackage{subfigure}
\usepackage{amsthm,amsmath,amssymb}
\usepackage{hyperref}
\journal{International Journal of Forecasting}

\begin{document}

\begin{frontmatter}

%% Title, authors and addresses

%% use the tnoteref command within \title for footnotes;
%% use the tnotetext command for theassociated footnote;
%% use the fnref command within \author or \address for footnotes;
%% use the fntext command for theassociated footnote;
%% use the corref command within \author for corresponding author footnotes;
%% use the cortext command for theassociated footnote;
%% use the ead command for the email address,
%% and the form \ead[url] for the home page:
%% \title{Title\tnoteref{label1}}
%% \tnotetext[label1]{}
%% \author{Name\corref{cor1}\fnref{label2}}
%% \ead{email address}
%% \ead[url]{home page}
%% \fntext[label2]{}
%% \cortext[cor1]{}
%% \affiliation{organization={},
%%             addressline={},
%%             city={},
%%             postcode={},
%%             state={},
%%             country={}}
%% \fntext[label3]{}

\title{Wind energy forecasting with missing values within a fully conditional specification framework}

\author[inst1,inst3]{Honglin Wen}
\author[inst2,inst3]{Pierre Pinson}
\author[inst1]{Jie Gu}
\author[inst1]{Zhijian Jin}

\affiliation[inst1]{organization={Department of Electrical Engineering, Shanghai Jiao Tong University}}

\affiliation[inst2]{organization={Dyson School of Design Engineering, Imperial College London}}

\affiliation[inst3]{organization={Department of Technology, Management and Economics, Technical University of Denmark}}

\begin{abstract}
Wind power forecasting is essential to power system operation and electricity markets. As abundant data became available thanks to the deployment of measurement infrastructures and the democratization of meteorological modeling, extensive data-driven approaches have been developed within both point and probabilistic forecasting frameworks. These models usually assume that the dataset at hand is complete and overlook missing value issues that often occur in practice. In contrast to that common approach, we rigorously consider here the wind power forecasting problem in the presence of missing values, by jointly accommodating imputation and forecasting tasks. Our approach allows inferring the joint distribution of input features and target variables at the model estimation stage based on incomplete observations only. We place emphasis on a fully conditional specification method owing to its desirable properties, e.g., being assumption-free when it comes to these joint distributions. Then, at the operational forecasting stage, with available features at hand, one can issue forecasts by implicitly imputing all missing entries. The approach is applicable to both point and probabilistic forecasting, while yielding competitive forecast quality within both simulation and real-world case studies. It confirms that by using a powerful universal imputation method based on fully conditional specification, the proposed universal imputation approach is superior to the common impute-then-predict approach, especially in the context of probabilistic forecasting.

\end{abstract}

%%Graphical abstract
%\begin{graphicalabstract}
%\includegraphics{grabs}
%\end{graphicalabstract}

%%Research highlights
%\begin{highlights}
%\item Research highlight 1
%\item Research highlight 2
%\end{highlights}

\begin{keyword}
%% keywords here, in the form: keyword \sep keyword
Wind power  \sep Probabilistic forecasting \sep Missing values \sep Multiple imputation
%% PACS codes here, in the form: \PACS code \sep code
%\PACS 0000 \sep 1111
%% MSC codes here, in the form: \MSC code \sep code
%% or \MSC[2008] code \sep code (2000 is the default)
%\MSC 0000 \sep 1111
\end{keyword}

\end{frontmatter}

%% \linenumbers

%% main text
\section{Introduction}

\subsection{Background}

As a cornerstone to achieve net-zero emissions in the energy sector, wind power has proliferated over recent decades. However, the stochastic nature of wind power generation challenges power system operation and electricity markets, which has therefore motivated wind power forecasting~(WPF) research. WPF is usually classified into short-term forecasting (hours to a few days) which takes numerical weather predictions as input features and very short-term forecasting (minutes to a few hours) which utilizes recent observations as input features. Recently WPF has achieved several advances by employing cutting-edge statistical and machine learning approaches, e.g. deep learning \citep{Goodfellow-et-al-2016} and lightGBM \citep{ke2017lightgbm}, as well as the modeling of underlying stochastic processes through the investigation of its spatial-temporal dynamics \citep{cavalcante2017lasso, messner2019online}.

Meanwhile, the interest of the WPF community has shifted from point forecasting to probabilistic forecasting; see recent review by \citet{hong2020energy}.
Probabilistic wind power forecasting~(PWPF) communicates the probability distribution of wind power generation at a future time based on gathered information up to the issue time, usually in the form of quantiles, prediction intervals, and densities.
It has attracted increasing attention in the power industry, especially after the 2014 Global Energy Forecasting Competition~(GEFCom 2014) \citep{Hong2016Probabilistic}.
In general, two approaches, namely parametric and non-parametric have been proposed for PWPF.
The parametric approach is based on a distributional assumption, such as Gaussian, Beta, etc., the shape parameters of which are determined through statistical learning.
In contrast, the non-parametric approach is free of such an assumption.
One of the most popular non-parametric approaches relies on quantile regression~(QR) \citep{koenker2001quantile}, which involves a pinball loss function to guide the learning of conditional quantile functions. It is therefore easy to employ QR in advanced statistical learning models (for instance gradient boosting machine \citep{LANDRY20161061} and extreme learning machine \citep{wan2016direct}) by using the pinball loss as loss function at the model estimation phase. Besides, with the aim to characterize the whole distribution in a distribution-free manner, methods that simultaneously estimate several quantiles \citep{sangnier2016joint} and directly estimate the distribution based on conditional normalizing flow(s) \citep{wen2022continuous} have been proposed.

Although several works have contributed forecasting methods and products to the WPF community, most of them assume that the dataset at hand is complete and overlook the widespread missing value problems, due to sensor failures and communication errors for instance.
Intuitively, missing value issues pose problems at both model estimation and operational forecasting stages, ultimately compromising forecast quality.
Obviously, for models estimated through gradient-based optimization, the training datasets cannot contain missing values, otherwise, the gradients of the parameters cannot be calculated at the model estimation stage.
Therefore, rows of the learning set containing both missing values and observations are often deleted, even if the missing information is minimal. It means that valuable information is also discarded in the process of removing the missing values. In addition, even with estimated models at hand, missing value problems still affect operational forecasting, possibly obliging forecasters to revert to naive models such as climatology (i.e., long-term averages) as surrogates.
Therefore, it remains an open issue to investigate the influence of missing values and develop WPF approaches that accommodate missing values.

\subsection{Related works}

An intuitive and popular approach to the problem (though, not used by the WPF community) is to impute these missing values before training models and issuing operational forecasts  \citep{liu2018wind}. It is referred to as \emph{``impute, then predict'' (ITP)} approach in this paper. For example, the classic forecasting package ``forecast'' \citep{hyndman2008automatic} provides an option that uses linear interpolation to impute missing values. Obviously, a spectrum of imputation methods can be employed; see a thorough review by \citet{van2018flexible}. Then, an associated question is how to choose the imputation method. The recent study by \citet{tawn2020missing} suggests that the influence of imputation on model estimation and operational forecasting stages is ambiguous. Concretely, they concluded that advanced imputation methods are beneficial to model estimation. However, at the operational forecasting, it turns out that retraining models without missing features results in better performance. In fact, it is natural to consider the retraining approach, as it only uses actual observations to estimate parameters, and hence prevents using the aforementioned imputation procedure. However, the learning then only relies on a subset of the data available, while the information potentially contained in the discarded part is lost. In addition, this approach may suffer the curse of dimensionality since having to train models for all combinations of input features. This may yield a substantial increase in computational costs.

In addition to the aforementioned approaches, several works have focused on adapting forecasting methods to be used in the presence of missing values. A classic approach is based on state-space modeling, where the Kalman filter is modified to allow accommodating incomplete observations. For example, autoregressive moving average models \citep{jones1980maximum} and autoregressive integrated moving average models \citep{kohn1986estimation} have been represented in state-space form and adapted to tackle missing value problems. Although these works have shed light on forecasting in the presence of missing values, they are only applicable to point forecasting and restricted to linear models. Recent advanced models such as GRU-D \citep{che2018recurrent} and BRITS \citep{cao2018brits} have been proposed based on the long-short term memory model \citep{hochreiter1997long}, by using the intermediate results (which can be also interpreted as latent states) of the neural network model to impute missing values. This idea has been successfully applied in the recent popular package DeepAR \citep{salinas2020deepar}. However, they still require imputing missing values via the recurrent neural network structure before performing the forecasting task.

\subsection{Proposed method and contributions}

There is no such a clearly defined boundary between imputation and forecasting, as explained by \citet{golyandina2007caterpillar}. Indeed, a forecasting problem can be considered as an imputation problem in the situation where missing values are systematically located at the end of a sequence. Furthermore, both imputation and forecasting tasks assume the continuation of the underlying structure within the data, and consequently leverage observations to predict unknown values. That is, it is feasible to develop a model that can infer the structure based on observations and seamlessly perform the imputation and forecasting tasks, which is referred to as \emph{``universal imputation''} (UI) approach in this paper. As a result, in what follows, we may interchangeably use the terms ``impute'' and ``forecast''. With this idea in mind, \citet{NEURIPS2020_dc36f18a} considered the point forecasting problem and proposed to model the correlation structure between input features and targets via a graph neural network, where imputation of missing features and prediction of targets can be simultaneously performed. In contrast here, we place ourselves within a probabilistic setting directly, for which it is then also possible to derive point forecasts. Unlike the usual probabilistic forecasting approaches that model conditional probability distributions (for the target variable) directly, in this work we model the joint multivariate probability distribution of input features and targets. As discussed by \citet{stone1991asymptotics}, with the estimated multivariate probability distribution at hand, one can obtain conditional distributions via marginalization, although it is computationally inefficient compared to the usual conditional probability distribution modeling. This approach is appealing in the presence of missing values. That is, with the estimated multivariate probability distribution, one can marginalize over missing variables to obtain probabilistic forecasts. Then, the goal at the model estimation stage is to estimate the parameters of such a distribution based on incomplete observations. At the operational forecasting stage, targets to be predicted are treated as missing values and imputed via the estimated distribution. The focus is on a very-short-term wind power forecasting applications, where missing value issues often occur, though the method is generic and could be then used by others for different applications where challenges accommodating missing values are also present. 

In this work, we focus on situations where observations are missing at random due to, e.g., sensor failures and communication errors. This means missingness patterns are independent of the missing values themselves. However, it does not mean that this concept of missingness at random is restricted to the case of data missing in a pointwise and sporadic fashion. Even in the case of block missingness (i.e., data missing over time intervals), as long as the data is missing at random (hence, independently of the values for the process of interest or relevant exogenous processes), our approach can be employed. The distribution of missingness can then be left aside when inferring the underlying structure of interest. As a consequence, the problem boils down to estimating the parameters of a model based on incomplete observations only. Missing-not-at-random cases could still be handled by the proposed UI approach, though requiring more sophisticated techniques at both the model estimation and operational forecasting stages, which is left for future work. Specifically, it requires modeling the missingness explicitly when calculating the likelihood at the model estimation stage. At the operational forecasting stage, the targets to be predicted are treated as missing, and thus independent of the missingness distribution of the contextual features. It also requires taking into account the missingness distribution of contextual features when calculating the conditional distribution of missing variables given the observed variables. Particularly, we implement this idea based on the multiple imputation method \citep{dempster1977maximum}, which allows us to impute missing values with several equally likely realizations from the distribution and thus provides probabilistic forecasts for the targets. Instead of assuming a special family of distributions and inferring its parameters, we adopt the fully conditional specification~(FCS) approach \citep{van2006fully}, which implicitly specifies the multivariate distribution as a collection of conditional distributions on a variable-by-variable basis. At the model estimation stage, parameters for each conditional distribution are iteratively estimated through a Gibbs sampling procedure. At the operational forecasting stage, missing values are also iteratively imputed on a variable-by-variable basis.

The proposed method is validated based on a simulation study and real-world case studies with wind power data from the USA. The simulation study is based on synthetic data (for both AR and VAR processes) and Monte-Carlo simulations, to illustrate and underline the salient features of our approach. It also allows analyzing the impact of certain characteristics e.g. rate of missingness on the performance of the approach, while remaining in a controlled environment within which the results are due to changes in the design, and not to some spurious effects often observed with real-world data. Real-world data from the US are used for benchmarking instead, and to investigate various aspects of the applicability of the approach in an operational context, e.g., with a focus on pointwise vs. block missingness, various rates of missingness, univariate and multivariate setups, etc. The results show that the proposed approach is superior to existing ITP approaches. The main contributions of this paper are two-fold. One of them is the proposal of a universal imputation approach, which is general, though inspired by the problem of wind power forecasting in the presence of missing values. Such a universal imputation approach jointly accommodates imputation and forecasting tasks within the universal multiple imputation framework. By design, this approach allows to generate both point and probabilistic forecasts. The other contribution is to show its applicability to wind power forecasting with missing values, where different types and rates of missingness are present.

The remaining parts of this paper are organized as follows.
Section 2 formulates the problem, whereas Section 3 describes the proposed approach for forecasting in the presence of missing values.
Next, the simulation study to show the applicability of the proposed approach is elaborated in Section 4.
Section 5 presents case studies with results and discussion.
Section 6 concludes the paper.

\vspace{1em}
\emph{Notations:} In general, we use uppercase letters to denote random variables and lowercase letters to denote the realizations of these random variables.
For instance, $Y_1$ denotes a random variable and $y_1$ its realization.
A collection of random variables are represented as a tuple, which is bracketed with parentheses, such as $(Y_1, Y_2)$ and $(Y_1,\cdots, Y_{10})$.
Boldface lowercase and uppercase letters respectively indicate vectors and matrices.
Particularly, we use row and column slices to represent parts of a matrix.
For instance, let $\boldsymbol Z$ represent a matrix, $\mathcal{I}$, and $\mathcal{J}$ denote row indices and column indices.
Then, $\boldsymbol Z[\mathcal{I};\mathcal{J}]$ represents a part of matrix $\boldsymbol Z$ indexed by $\mathcal{I}$ and $\mathcal{J}$.
And, $(\cdot)^\top$ denotes the transpose of matrices.
A time series is represented as $\{y_t,t=1,2,\cdots \}$ indexed by time $t$, which is a realization of a stochastic process $\{Y_t, t=1,2,\cdots \}$.
We also write them as $\{y_t \}$ and $\{Y_t \}$ for short.

\section{Preliminaries}

We first describe the framework for very-short term wind power forecasting, for both point and probabilistic forecasting cases. Subsequently, we detail the challenges induced by missing values at both model estimation and operational forecasting stages.

\subsection{Problem Formulation}
Assume we have $p$ wind farms in a region that can share information to improve forecasting accuracy as suggested by \citet{cavalcante2017lasso}.
When $p$ equals to 1, it reduces to the common single wind farm case.
At wind farm $n$, let $y_{n,t}\in [0,P_n]$ (where $P_n$ is its capacity) denote the wind power generation value at time $t$, which is a realization of the random variable $Y_{n,t}$.
Let $\Omega_{n,t}$ denote the information tuple of wind farm $n$ up to time $t$, which would contain values over previous time steps and possibly other relevant information such as weather observations and numerical weather forecasts.
And let $\Omega_t$ represent the tuple that contains information of all sites up to time $t$, i.e., $\Omega_t=( \Omega_{1,t},\cdots,\Omega_{p,t})$.
Generally, the aim is to issue forecasts with lead time $h$, i.e., the characteristics of $Y_{1,t+1},\cdots,Y_{1,t+h},\cdots,Y_{p,t+1},\cdots,Y_{p,t+h}$, given information $\Omega_t$.
The forecasting task can be decoupled into several sub-problems, each of which focuses on a specific site and time, for instance forecasting the characteristics of $Y_{n,t+h}$ based on the whole information pool $\Omega_t$.
Then the point forecast for $Y_{n,t+h}$ given by a model $\mathcal{M}$ with parameters $\hat{\Theta}_t$ is usually defined as
\begin{equation}
    \hat{y}_{n,t+h|t}=\mathbb{E}[Y_{n,t+h}|\mathcal{M},\hat{\Theta}_t,\Omega_t],
\end{equation}
where $\mathbb{E}[\cdot]$ denotes the expectation of random variables, and $\hat{\Theta}_t$ changes with time $t$.
In this paper, let us assume the stochastic process $\{ Y_{1,t}, Y_{2,t},\cdots,Y_{p,t}\}$ is stationary.
Then, the density function $f_{Y_{1,t},\cdots,Y_{p,t+h}}$ is invariant for changes in time \citep{de2017elements}, which means parameters $\hat{\Theta}_t$ do not vary with time and are denoted as $\hat{\Theta}$.
Then, one can estimate the parameters based on collected data via statistical learning methods.
We rewrite (1) as
\begin{equation}
    \hat{y}_{n,t+h|t}=\mathbb{E}[Y_{n,t+h}|\mathcal{M},\hat{\Theta},\Omega_t].
\end{equation}
The probabilistic forecast for time $t+h$ given by $\mathcal{M}$ is communicated as a density function, i.e.,
\begin{equation} \label{eq:preddens}
    \hat{f}_{n,t+h|t}(y)=f_{Y_{n,t+h}}(y|\mathcal{M},\hat{\Theta},\Omega_t).
\end{equation}
Indeed, with the estimated density function at hand, one can easily obtain point forecast via:
\begin{equation}
    \hat{y}_{n,t+h|t}=\int_y y \, \hat{f}_{n,t+h|t}(y)\ d y.
\end{equation}
For simplicity of notations, let us focus on the predictive density $\hat{f}_{n,t+h|t}$ given the information set $\Omega_t$ at time $t$. We denote the input features as $\boldsymbol x_t$ and the realization of target $Y_{n,t+h}$ as $y_t$. The information  one has access to $N$ sample pairs $(\boldsymbol x_1,y_1),\cdots,(\boldsymbol x_N,y_N)$ serves as a basis for training. They can be written in the form of a matrix, i.e., $\boldsymbol X=[\boldsymbol x_1,\cdots,\boldsymbol x_N ]^\top$ as well as $\boldsymbol Y=[ y_1,\cdots, y_N ]^\top$. The matrix $\boldsymbol X$ has dimensions $N\times pk$, whereas $\boldsymbol Y$ is a vector with $N$ elements.
Now the density forecast described in~\eqref{eq:preddens} boils down to conditional probability density function estimation, which is performed via statistical learning.
In very-short term WPF, one commonly uses past wind power generation values of length $k$ as input features, i.e., a vector $[y_{n,t-k+1},\cdots,y_{n,t}]^\top \in [0,P_n]^k$ for the $n^\text{th}$ site. Therefore, considering all sites together, the vector of input features is given by
\begin{align*}
    \boldsymbol x_t=[ y_{1,t-k+1},\cdots,y_{1,t},\cdots, y_{p,t-k+1},\cdots, y_{p,t} ]^\top \in [0,P_1]^{k}\times[0,P_2]^{k}\times \cdots \times [0,P_p]^{k}.
\end{align*}
%Assume the time series $\{y_{1,t},y_{2,t},\cdots,y_{p,t}\}$ up to time T is available.
Obviously, features in $\boldsymbol{x}_t$ have some form of dependency, which breaks down the classical i.i.d assumption in statistical learning. However, it is still common to place oneself in a regression framework for estimation and overlook this dependency issue, as for instance done recently also for global/local model estimation \citep{MONTEROMANSO20211632} and estimation in deep learning models \citep{2022deep}. The consequences are actually fairly mild in practice, since the fact that input features are not independent mainly affects the interpretability of regression coefficients and the ability to perform hypothesis testing (to assess whether coefficients are significantly different from 0). This is while the fact that observation samples used as a basis for estimation are not i.i.d. mainly yields higher variance in coefficient estimates -- an issue if dealing with small datasets, which is rarely the case today for most statistical and machine learning applications, including wind energy forecasting.

Based on a stationarity assumption,  the sample pairs $(\boldsymbol x_1,y_1),\cdots,(\boldsymbol x_N,y_N)$ can be regarded as identically distributed.
For simplicity, we introduce two random variables $X$ and $Y$ for these samples.
It allows us to model the joint distribution $f_{X,Y}(\boldsymbol x,y)$ via $\mathcal{M}$ with estimated parameters $\hat{\Theta}$, i.e., $f_{X,Y}(\boldsymbol x,y;\mathcal{M},\hat{\Theta})$, and derive $f_{Y|X}(y|\boldsymbol x)$ via the conditional probability formula.
It is described as
\begin{equation}
    f_{Y|X}(y|\boldsymbol x;\mathcal{M},\hat{\Theta})=\frac{f_{X,Y}(\boldsymbol x,y;\mathcal{M},\hat{\Theta})}{f_X(\boldsymbol x;\mathcal{M},\hat{\Theta})}=\frac{f_{X,Y}(\boldsymbol x,y;\mathcal{M},\hat{\Theta})}{\int_y f_{X,Y}(\boldsymbol x,y;\mathcal{M},\hat{\Theta}) dy}.
\end{equation}
With the estimated joint distribution $f_{X,Y}(\boldsymbol x,y;\mathcal{M},\hat{\Theta})$ at hand, at any time $t$, given contextual information $\boldsymbol x_t$, one can issue the forecast $\hat{f}_{Y|X}(y_t|\boldsymbol x_t;\mathcal{M},\hat{\Theta})$ via (5).
In this paper, $\mathcal{M}$ is set as an imputation model and implicitly defined by a collection of conditional distributions.
Each conditional distribution is implemented by predictive mean matching that relies on a function, for instance $g_j$ parameterized by $\hat{\theta}_j$.
As we are considering the joint distribution now, we can concatenate $\boldsymbol x_t$ and $y_t$ as $\boldsymbol z_t$, i.e., $\boldsymbol z_t=[\boldsymbol x_t^\top, y_t]^\top$.
Accordingly, the dataset is concatenated as the matrix $\boldsymbol Z$ of shape $N\times (pk+1)$, i.e.,
\begin{center}
    $\boldsymbol Z =
    \begin{bmatrix}
    \boldsymbol x_1^\top & y_1\\
    \boldsymbol x_2^\top & y_2\\
    \vdots & \vdots \\
    \boldsymbol x_N^\top & y_N
    \end{bmatrix}=
    \begin{bmatrix}
    \boldsymbol z_1^\top \\
    \boldsymbol z_2^\top \\
    \vdots \\
    \boldsymbol z_N^\top
    \end{bmatrix}$.
\end{center}
We refer to the $i$-th row, $j$-th column, and $(i,j)$-th entry of $\boldsymbol Z$ as $\boldsymbol z_i$, $\boldsymbol Z_j$, and $z_{i,j}$ respectively.
And we introduce a random variable $Z=(X,Y)$ that concatenates $X$ and $Y$, which contains $pk+1$ variables (recall that $X$ has $pk$ variables, as it represents information from $p$ sites), i.e., $Z=(Z_1,Z_2,\cdots,Z_{pk+1})$.
Then, the distribution of $Z$ is modeled by $f_Z(\boldsymbol z;\mathcal{M},\hat{\Theta})$.
In particular, let $Z_{-j}$ denote the collection of random variables in $Z$ except $Z_j$, i.e., $Z_{-j}=(Z_1,\cdots,Z_{j-1},Z_{j+1},\cdots,Z_{pk+1} )$. 
Accordingly, let $\boldsymbol z_{-j}$ denote the realization of $Z_{-j}$.

We assume values are missing at random. This is to be understood in a way that is more general than data missing sporadically and at random times. More formally, missingness at random means that the fact a data entry is missing or not is independent of the process itself, or of some exogenous process. For the wind power application, data missing not at random could be for the case there are systematic sensor failures for power generation values below a given threshold, or systematic communication failures when wind comes for a given direction. In addition, missingness at random is not restricted to the case data is missing at single times. It can also be for the case of data missing over time intervals (i.e., block missingness). This assumption of missingness at random is expected to be sound for wind power applications, though this should be confirmed on a case-by-case basis based on advanced data analysis. 

\begin{figure}[!ht]
\centering
\includegraphics[width=1.75in]{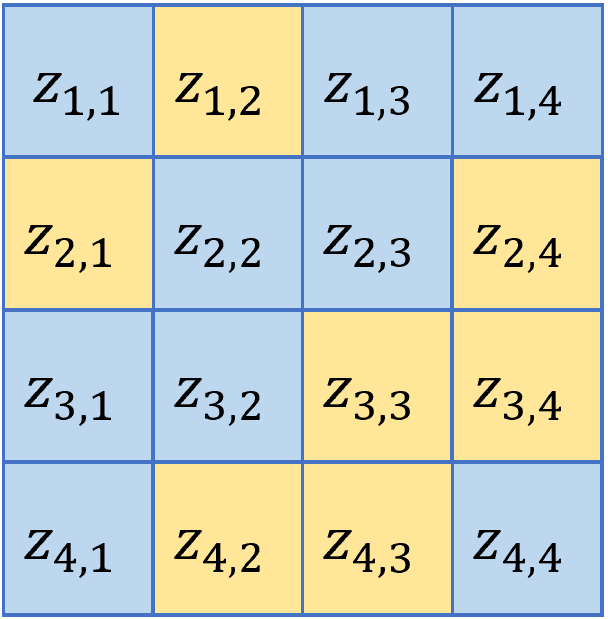}
\caption{Illustration of a dataset $\boldsymbol Z$. Here we take $p=1$, $k=3$, $h=1$ as an example. Blue blocks indicate observations, whereas yellow blocks indicate missing values.}
\label{data z}
\end{figure}

Missing values are likely to occur in every element of $\boldsymbol z_t$.
Let us introduce a vector $\boldsymbol m_t$ to indicate the missingness of $\boldsymbol z_t$.
Concretely, $m_{t,j}=1$ indicates that $z_{t,j}$ is missing, whereas $m_{t,j}=0$ indicates that $z_{t,j}$ is observed.
Accordingly, the matrix $\boldsymbol M$ indicates the missingness of $\boldsymbol Z$. 
Let $\mathcal{J}_{\boldsymbol z_t,M}$ denote the indices of missingness of $\boldsymbol z_t$, i.e., $\mathcal{J}_{\boldsymbol z_t,M}=\{j\ |\ m_{t,j}=1 \}$, and $\mathcal{J}_{\boldsymbol z_t,O}$ denote the indices of observations, i.e., $\mathcal{J}_{\boldsymbol z_t,O}=\{j\ |\ m_{t,j}=0 \}$.
Therefore, the observed and missing parts of $\boldsymbol z_t$ are represented by $\boldsymbol z_t[\mathcal{J}_{\boldsymbol z_t,O}]$ and $\boldsymbol z_t[\mathcal{J}_{\boldsymbol z_t,M}]$, which are written as $\boldsymbol z_t^{obs}$ and $\boldsymbol z_t^{mis}$ for simplicity.
The corresponding random variables for $\boldsymbol z_t^{obs}$ and $\boldsymbol z_t^{mis}$ are denoted as $Z^{obs}$ and $Z^{mis}$.
When $y_t$ is missing, $\boldsymbol z_t^{mis}=[ {\boldsymbol x_t^{mis}}^\top,y_t]^\top$ where $\boldsymbol x_t^{mis}$ is the missing part of $\boldsymbol x_t$.
The corresponding random variables for $\boldsymbol x_t^{mis}$ are denoted as $X^{mis}$. 
For example, Figure~\ref{data z} presents the matrix $\boldsymbol Z=[z_{i,j}]_{4\times 4}$, where blue blocks indicate observations and yellow blocks indicate missing values.
As shown, the first row of $\boldsymbol Z$ is denoted as $\boldsymbol z_1$, the second entry of which, i.e., $z_{1,2}$ is missing.
Then, the indices of missing values and observations of $\boldsymbol z_1$ are $\mathcal{J}_{\boldsymbol z_1,M}=\{2\}$ and $\mathcal{J}_{\boldsymbol z_1,O}=\{1,3,4\}$.
Accordingly, we have $\boldsymbol z_1^{obs}=[z_{1,1},z_{1,3},z_{1,4}]^\top$, $\boldsymbol z_1^{mis}=[z_{1,2}]$.
The corresponding random variables for  $\boldsymbol z_1^{obs}$ and $\boldsymbol z_1^{mis}$ are denoted as $Z^{obs}=(Z_1,Z_3,Z_4)$ and $Z^{mis}=Z_2$.
Also, let $\mathcal{I}_{\boldsymbol Z_j,M}$ denote the indices of missing values in $\boldsymbol Z_j$, i.e., $\mathcal{I}_{\boldsymbol Z_j,M} = \{i\ |\ m_{i,j}=1\}$, and $\mathcal{I}_{\boldsymbol Z_j,O}$ denote the indices of observations in $\boldsymbol Z_j$, i.e., $\mathcal{I}_{\boldsymbol Z_j,O} = \{i\ |\ m_{i,j}=0\}$.
Then the missing and observed parts of $\boldsymbol Z_j$ are $\boldsymbol Z[\mathcal{I}_{\boldsymbol Z_j,M};j]$ and $\boldsymbol Z[\mathcal{I}_{\boldsymbol Z_j,O};j]$, which are respectively written as $\boldsymbol Z_{j}^{mis}$ and $\boldsymbol Z_{j}^{obs}$ for simplicity.
In Figure~\ref{data z}, $\boldsymbol Z_1$ represents the first column of $\boldsymbol Z$, the second entry of which is missing.
Accordingly, we have $\mathcal{I}_{\boldsymbol Z_1,M}=\{2\}$, $\mathcal{I}_{\boldsymbol Z_1,O}=\{1,3,4\}$, $\boldsymbol Z_1^{obs}=[z_{1,1},z_{3,1},z_{4,1}]^\top$, and $\boldsymbol Z_1^{mis}=[z_{2,1}]$.

\begin{figure}[!ht]
\centering
\includegraphics[width=3.4in]{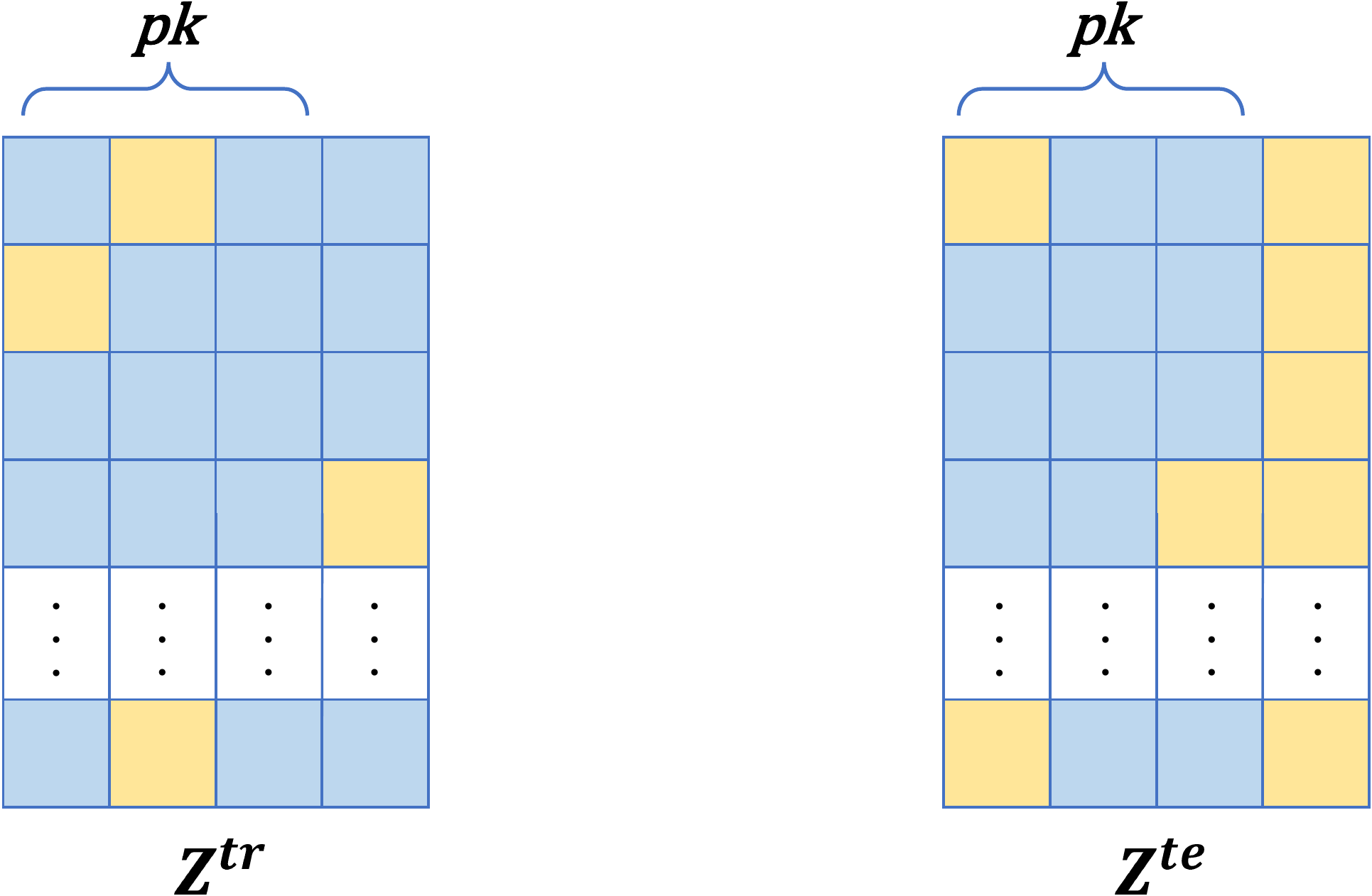}
\caption{Illustration of training and test datasets. Here we take $p=1$, $k=3$, $h=1$ as an example. Blue blocks indicate observations, whereas yellow blocks indicate missing values.}
\label{data set}
\end{figure}

Therefore, at the model estimation phase, we concatenate features and targets to form a training dataset $\boldsymbol Z^{tr}$ with some missing values, based on which an imputation model $\mathcal{M}$ is trained.
At the operational forecasting phase, the input vector $\boldsymbol x_t$ at time $t$ is available, part of which may be missing, and we focus on target $y_t$.
Together, they form $\boldsymbol z_t=[\boldsymbol x_t^\top,y_t]^\top$.
Then $\boldsymbol z_t$ is imputed via the estimated model.
For illustration, we present the training and test datasets for the single wind farm case in Figure~\ref{data set}.
In the training dataset, missingness occurs in both input features and targets.
In the test dataset, all targets are systematically missing.

\subsection{Challenge at the model estimation stage}

Usually, the learning process of parameters in density estimation problems is based on maximum likelihood, which involves the computation of likelihood.
However, in the presence of missing values, the likelihood is blended with missingness indicators.
With the assumption that values are missing at random, the parameters of underlying distributions can be estimated based on observations only.
Let $\Theta$ denote the true parameters of $\mathcal{M}$.
Consider the likelihood of a sample $\boldsymbol z_t$.
It is described as
\begin{equation}
    f_Z(\boldsymbol z_t,\boldsymbol m_t;\mathcal{M},\Theta)=f_Z(\boldsymbol z_t^{obs},\boldsymbol z_t^{mis},\boldsymbol m_t;\mathcal{M},\Theta),
\end{equation}
where $\boldsymbol z_t^{mis}$ is missing.
The likelihood function can be marginalized with respect to $\boldsymbol z_t^{mis}$, i.e.,
\begin{equation}
\begin{split}
    f_{Z^{obs}}(\boldsymbol z_t^{obs};\mathcal{M},\Theta)&=\int f_{Z^{obs},Z^{mis}}(\boldsymbol z_t^{obs},\boldsymbol z_t^{mis},\boldsymbol m_t;\mathcal{M},\Theta)d \boldsymbol z_t^{mis}\\
    &=\int f_{Z^{obs},Z^{mis}}(\boldsymbol z_t^{obs},\boldsymbol z_t^{mis};\mathcal{M},\Theta)d \boldsymbol z_t^{mis}.
\end{split}
\end{equation}
Therefore, to learn the parameters $\Theta$, it is required to maximize the likelihood of observations only, i.e., $f_{Z^{obs}}(\boldsymbol z_i^{obs};\mathcal{M},\Theta)$.
The estimate of $\Theta$ is denoted as $\hat{\Theta}$.

\begin{comment}
    \vspace{3mm}
\textcolor{blue}{\textit{Remark:} If the missingness-at-random assumption cannot be deemed valid, the missingness patterns are required to be modeled explicitly. Let us denote the distribution of $\boldsymbol{m}_t$ as $f_M(\boldsymbol{m}_t;\phi)$, where $\phi$ represents parameters. Then the likelihood is calculated as
\begin{align*}
    f_{Z,M}(\boldsymbol z_t,\boldsymbol m_t;\mathcal{M},\Theta,\phi)=\int f_Z(\boldsymbol z_t^{obs},\boldsymbol z_t^{mis};\mathcal{M},\Theta)f_M(\boldsymbol{m}_t|\boldsymbol z_t^{obs},\boldsymbol z_t^{mis};\phi)d \boldsymbol z_t^{mis}.
\end{align*}
That is, the parameters of both data and missingness distributions are jointly estimated.}
\end{comment}

\subsection{Challenge at the operational forecasting stage}

In this section, we assume that we already have distribution $f_Z(\boldsymbol z;\mathcal{M},\hat{\Theta})$ with estimated parameters $\hat{\Theta}$ at hand, and show how to issue forecasts at the operational forecasting stage.
If $\boldsymbol x_t$ is fully observed, then $\boldsymbol x_t$ is the observed part of $\boldsymbol z_t$, i.e., $\boldsymbol z_t^{obs}=\boldsymbol x_t$, whereas the missing part of $\boldsymbol z_t$ is $y_t$.
The forecast for $y_t$ can be expressed as
\begin{equation}
    f_{Y|X}(y_t|\boldsymbol x_t;\mathcal{M},\hat{\Theta})=f_{Z^{mis}|Z^{obs}}(\boldsymbol z_t^{mis}|\boldsymbol z_t^{obs};\mathcal{M},\hat{\Theta})=\frac{f_Z(\boldsymbol z_t^{obs},\boldsymbol z_t^{mis};\mathcal{M},\hat{\Theta})}{\int_{\boldsymbol z_t^{mis}} f_Z(\boldsymbol z_t^{obs},\boldsymbol z_t^{mis};\mathcal{M},\hat{\Theta}) d \boldsymbol z_t^{mis}}.
\end{equation}
In the presence of missing values, the forecasting task is to issue $f_{Y|Z^{obs}}(y|\boldsymbol z_t^{obs})$ by utilizing the distribution $f_Z(\boldsymbol z;\mathcal{M},\hat{\Theta})$.
Indeed, $\boldsymbol z_t^{mis} $ can be decomposed into $\boldsymbol x_t^{mis}$ and $y_t$, i.e.,
\begin{equation}
    f_{Z^{mis}|Z^{obs}}(\boldsymbol z_t^{mis}|\boldsymbol z_t^{obs};\mathcal{M},\hat{\Theta})=f_{Y,X^{mis}|Z^{obs}}(y_t,\boldsymbol x_t^{mis}|\boldsymbol z_t^{obs};\mathcal{M},\hat{\Theta}).
\end{equation}
Then the desired $f_{Y|Z^{obs}}(y_t|\boldsymbol z_t^{obs};\mathcal{M},\hat{\Theta})$ is derived by marginalizing $f_{Z^{mis}|Z^{obs}}(\boldsymbol z_t^{mis}|\boldsymbol z_t^{obs};\mathcal{M},\hat{\Theta})$ with respect to $\boldsymbol x_t^{mis}$, i.e.,
\begin{equation}
    f_{Y|Z^{obs}}(y_t|\boldsymbol z_t^{obs};\mathcal{M},\hat{\Theta})=\int f_{Y,X^{mis}|Z^{obs}}(y_t,\boldsymbol x_t^{mis}|\boldsymbol z_t^{obs};\mathcal{M},\hat{\Theta}) d \boldsymbol x_t^{mis}.
\end{equation}

\begin{comment}
    \vspace{3mm}
\textcolor{blue}{\textit{Remark:} If the missingness-at-random assumption cannot be deemed valid, it is required to take the distribution of missingness into account when calculating $f_{Z^{mis}|Z^{obs}}(\boldsymbol z_t^{mis}|\boldsymbol z_t^{obs};\mathcal{M},\hat{\Theta})$, which turns out to be $f_{Z^{mis}|Z^{obs},M}(\boldsymbol z_t^{mis}|\boldsymbol z_t^{obs},\boldsymbol{m}_t;\mathcal{M},\hat{\Theta},\hat{\phi})$ instead. Then the conditional distribution is calculated as
\begin{align*}
    f_{Z^{mis}|Z^{obs},M}(\boldsymbol z_t^{mis}|\boldsymbol z_t^{obs},\boldsymbol{m}_t;\mathcal{M},\hat{\Theta},\hat{\phi})=\frac{f_{Z,M}(\boldsymbol z_t^{obs},\boldsymbol z_t^{mis}, \boldsymbol{m}_t;\mathcal{M},\hat{\Theta},\hat{\phi})}{\int_{\boldsymbol z_t^{mis}} f_Z(\boldsymbol z_t^{obs},\boldsymbol z_t^{mis};\mathcal{M},\hat{\Theta})f_M(\boldsymbol{m}_t|\boldsymbol z_t^{obs},\boldsymbol z_t^{mis};\hat{\phi}) d \boldsymbol z_t^{mis}}.
\end{align*}
Then we arrive at $f_{Y|Z^{obs},M}(y_t|\boldsymbol z_t^{obs},\boldsymbol{m}_t;\mathcal{M},\hat{\Theta},\hat{\phi})$ by marginalizing with respect to $\boldsymbol x_t^{mis}$.} 
\end{comment}

\section{Forecasting with missing values via FCS}

In this section, we develop a forecasting approach based on the proposed universal imputation strategy. For that, we employ the fully conditional specification approach, which in practice will be based on Gibbs sampling. It is described in the first part of the Section. This FCS approach requires a method to derive conditional distributions, which we describe in the second part. Eventually, it also relies on the choice for a regression model (random forests here), covered in the third part of the section. Finally, we will describe how the overall approach can be readily used for genuine forecasting with missing data.

\begin{figure*}[!ht]
\centering
\includegraphics[width=6in]{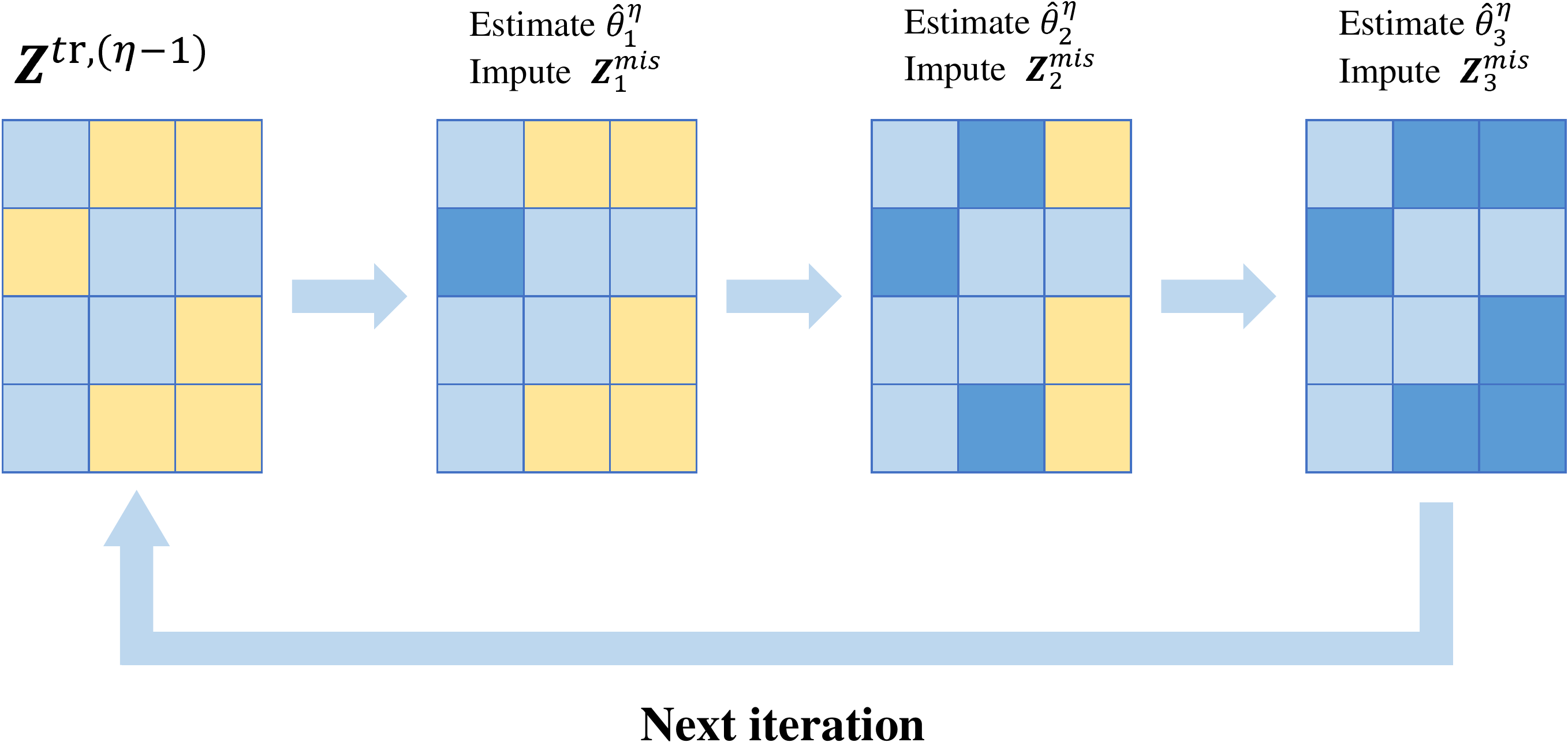}
\caption{Illustration of the $\eta$-th iteration at the training stage. Light blue blocks indicate observations, yellow blocks indicate missing values, and dark blue blocks indicate imputation.}
\label{training}
\end{figure*}

\subsection{Fully conditional specification method}

Instead of defining a multivariate distribution $f_Z(\boldsymbol z;\mathcal{M},\hat{\Theta})$ by assuming a specific distribution family, the FCS specifies a separate conditional distribution for each $Z_j$, just like a Gibbs sampler.
Concretely, the conditional distribution for $Z_j$ is modeled by $g_j$ with parameters $\hat{\theta}_j$, and is denoted as $f_{Z_j|Z_{-j}}( z_j|\boldsymbol z_{-j};g_j,\hat{\theta}_j)$.
Therefore, the model $\mathcal{M}$ is implemented via a bunch of models $\{ g_j\}$, whereas $\hat{\Theta}$ is composed of all parameters $\{ \hat{\theta}_j\}$.
These parameters are estimated at the model estimation phase based on the training dataset $\boldsymbol Z^{tr}$.
For simplicity of notations, we still use $\boldsymbol Z$ in what follows to show how to estimate the parameters.
Intuitively, before estimating $\hat{\theta}_j$, one needs to impute the missing values of $\boldsymbol Z_{-j}$.
Then, parameters are estimated based on the imputed $\boldsymbol Z_{-j}$ and $\boldsymbol Z_j^{obs}$.
With the estimated conditional distribution $f_{Z_j|Z_{-j}}( z_j|\boldsymbol z_{-j};g_j,\hat{\theta}_j)$, one can impute $\boldsymbol Z_{j}^{mis}$ based on the corresponding conditionals in $\boldsymbol Z_{-j}$.
That is, both the estimation of $\hat{\theta}_j$ and the imputation of $\boldsymbol Z_{j}^{mis}$ are based on the imputed $\boldsymbol Z_{-j}$.
Obviously, the imputation of any column of $\boldsymbol Z_{-j}$, for instance $\boldsymbol Z_{q}$, relies on its conditional distribution $f_{Z_{q}|Z_{-q}}( z_{q}|\boldsymbol z_{-q};g_{q},\hat{\theta}_{q})$, which requires $\boldsymbol Z_j^{mis}$ to be imputed.
In other words, the estimation of $\hat{\theta}_j$ and the imputation of $\boldsymbol Z_{-j}$ are coupled with each other.
If one performs the parameter estimation and imputation sequentially for $j=1,2,\cdots,pk+1$, the estimation of $\hat{\theta}_j$ can only use initial imputation of $\boldsymbol Z_{j+1}^{mis},\cdots,\boldsymbol Z_{pk+1}^{mis}$.
The updated imputation of $\boldsymbol Z_{j+1}^{mis},\cdots,\boldsymbol Z_{pk+1}^{mis}$ given by their estimated conditional distributions cannot be used for the estimation of $ \hat{\theta}_j$.
Therefore, we perform the imputation of $\boldsymbol Z_j$ and the estimation of $ \hat{\theta}_j$ in an iterative manner.
Then, at next iteration, the updated imputation of $\boldsymbol Z_{j+1}^{mis},\cdots,\boldsymbol Z_{pk+1}^{mis}$ can be used for the estimation of $ \hat{\theta}_j$.
For example, denote the estimated parameters $ \hat{\theta}_j$ at the $\eta$-th iteration as $\hat{\theta}_j^{(\eta)}$, and the imputed complete column as $\boldsymbol Z_j^{(\eta)}$.
At the $\eta+1$-th iteration, $\boldsymbol Z_{j+1}^{(\eta)},\cdots,\boldsymbol Z_{pk+1}^{(\eta)}$ can be used for the estimation of $\hat{\theta}_j^{(\eta+1)}$.
Before the iterative estimation, all missing values are initially imputed as 0; therefore each column $\boldsymbol Z_j$ becomes complete and is written as $\boldsymbol Z_{j}^{(0)}$.
After all iterations, the ultimate estimation for $\theta_j$ is denoted as $ \hat{\theta}_j$.
Here, we set the stopping criterion as the round of iteration, as suggested by \citep{van2006fully}.
The caveat is that FCS method cannot guarantee the existence of joint distribution.
Luckily, it is a relatively minor problem in practice, especially when missing rate is modest.
We illustrate the steps of the $\eta$-th iteration in Figure~\ref{training}.

Concretely, at the $\eta$-th iteration, before estimating $\hat{\theta}_j^{(\eta)}$, we have $\boldsymbol Z_1^{(\eta)}$,$\cdots$,$\boldsymbol Z_{j-1}^{(\eta)}$,$\boldsymbol Z_{j+1}^{(\eta-1)}$,$\cdots$, $\boldsymbol Z_{pk+1}^{(\eta-1)}$ at hand, which are written compactly as $\boldsymbol Z_{-j}^{(\eta)}$ in the form of a matrix, i.e.,
\begin{equation}
    \boldsymbol Z_{-j}^{(\eta)}=[\boldsymbol Z_1^{(\eta)},\cdots,\boldsymbol Z_{j-1}^{(\eta)},\boldsymbol Z_{j+1}^{(\eta-1)},\cdots,\boldsymbol Z_{pk+1}^{(\eta-1)}].
\end{equation}
Then $\hat{\theta}_j^{(\eta)}$ is estimated based on $\boldsymbol Z_{-j}^{(\eta)}$ and $\boldsymbol Z_{j}^{obs}$ via maximum likelihood:
\begin{equation}
    \hat{\theta}_j^{(\eta)}=\arg\max\limits_{\theta_j}\sum_{i\in \mathcal{I}_{j,obs}}\log f_{Z_j|Z_{-j}}(z_{i,j}|\boldsymbol z_{i,-j}^{(\eta)};g_j,\theta_j).
\end{equation}
Thus we derive the estimated conditional distribution $f_{Z_j|Z_{-j}}(z_j|\boldsymbol z_{-j};g_j,\hat{\theta}_j^{(\eta)})$, based on which we can impute $\boldsymbol Z_{j}^{mis}$.
For instance, to impute the value $z_{i,j}$ in $\boldsymbol Z_{j}^{mis}$, we sample from $f_{Z_j|Z_{-j}}(z_j|\boldsymbol z_{i,-j};g_j,\hat{\theta}_j^{(\eta)})$, which is described as:
\begin{equation}
    z_{i,j}^{(\eta)}\sim f_{Z_j|Z_{-j}}(z_j|\boldsymbol z_{-j};g_j,\hat{\theta}_j^{(\eta)}),\quad i \in \mathcal{I}_{j,mis}.
\end{equation}
As $\boldsymbol Z_{j}^{obs}$ is observed, we do not change the values, i.e.,
\begin{equation}
    z_{i,j}^{(\eta)}=z_{i,j}^{(\eta-1)},\quad i \in \mathcal{I}_{j,obs}.
\end{equation}
Then we write all $z_{i,j}^{(\eta)}$ in the form of a vector, which is denoted as $\boldsymbol Z_j^{(\eta)}$ i.e.,
\begin{equation}
\boldsymbol Z_j^{(\eta)}=[z_{1,j}^{(\eta)},\cdots, z_{N,j}^{(\eta)}]^\top.
\end{equation}
This procedure goes sequentially for $j=1,\cdots,pk+1$.
We note that the method can be executed multiple times in parallel to obtain multiple imputations.
Besides, the model $g_j$ for $f_{Z_j|Z_{-j}}( z_j|\boldsymbol z_{-j};g_j,\hat{\theta}_j)$ needs to be specified, which is described in next section.

\begin{figure}[!ht]
\centering  
\subfigure[Training stage]{   
\begin{minipage}{7cm}
\centering   
\includegraphics[width=2in,height=2.5in]{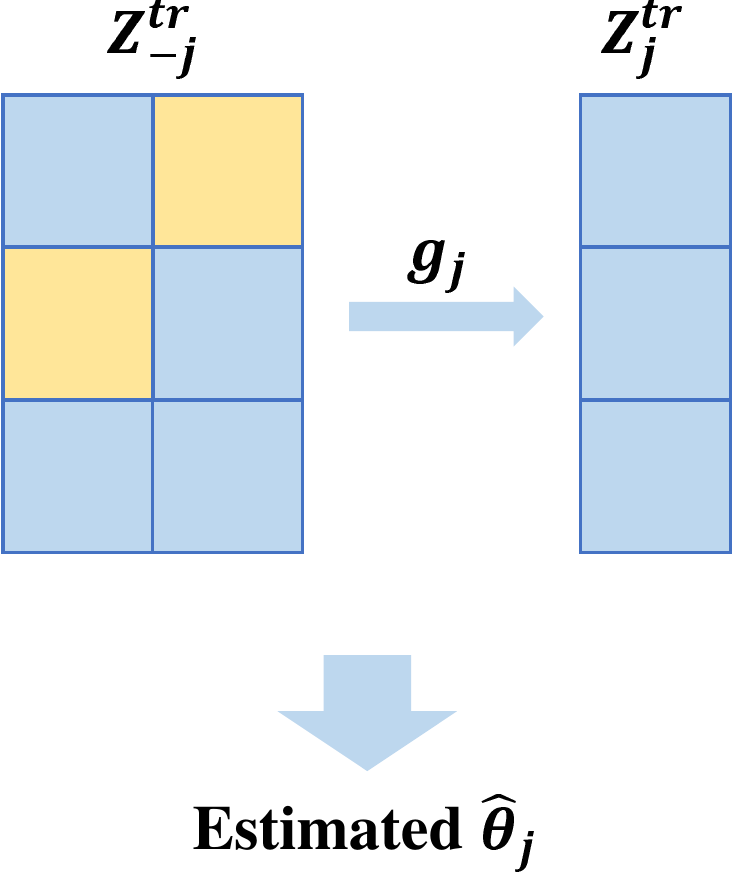}  
\end{minipage}
}
\subfigure[Candidates prediction]{ 
\begin{minipage}{7cm}
\centering   
\includegraphics[width=2in,height=2.5in]{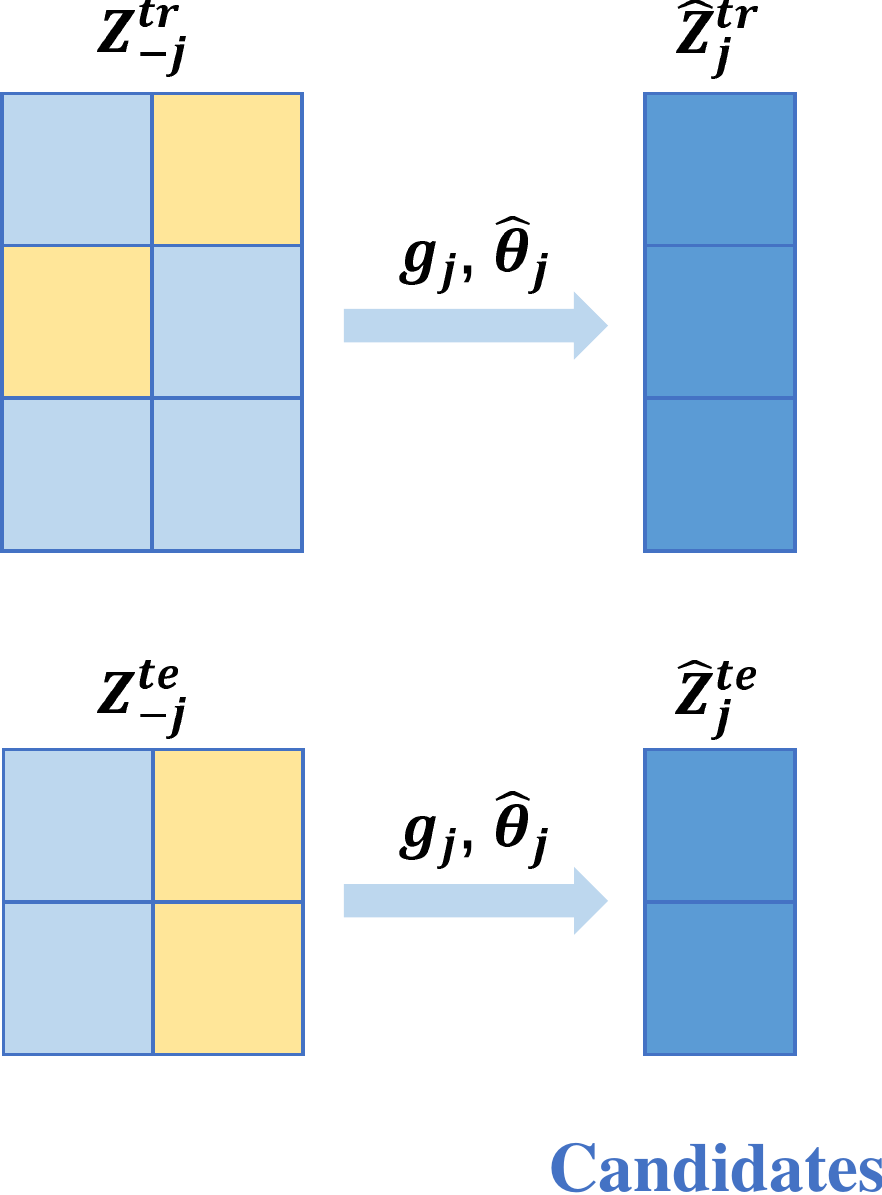}
\end{minipage}
}
\caption{Illustration of key components of predictive mean matching.}    
\label{mathcing}   
\end{figure}

\subsection{Predictive mean matching}

In this paper, $f_{Z_j|Z_{-j}}( z_j|\boldsymbol z_{-j};\hat{\theta}_j,g_j)$ is specified based on the predictive mean matching \citep{little2019statistical}, which is free of distributional assumptions.
Specifically, here $g_j$ is not a real distribution model, but specified as a regression model.
The distribution is given by a sampling procedure based on $g_j$.
For each missing entry, we form a set of candidates from complete cases whose predicted values are close to the predicted value for the missing entry.
Now we use parameters $\theta_j$ to specify the regression model $g_j$ that maps $\boldsymbol z_{-j}$ to $z_j$, i.e.,
\begin{equation}
    z_j = g_j(\boldsymbol z_{-j};\theta_j)+\epsilon_j,
\end{equation}
where $\epsilon_j$ represents noise.
We illustrate the key operations of this method in Figure~\ref{mathcing}, i.e., the training of the regression model and the prediction of candidates.
That is, we estimate parameters $\hat{\theta}_j$ based on training datasets $\boldsymbol Z_{-j}^{tr}$ and $\boldsymbol Z_j^{tr}$.
With the estimated model, we respectively predict targets for $\boldsymbol Z_{-j}^{tr}$ and $\boldsymbol Z_{-j}^{te}$, which are called candidates and written as $ \hat{\boldsymbol Z}_j^{tr}$ and $\hat{\boldsymbol Z}_j^{te}$.
Then, for each entry of $\hat{\boldsymbol Z}_j^{te}$, we form a set of its $d$ closest candidates in $\hat{\boldsymbol Z}_j^{tr}$, from which we perform random sampling to obtain imputations.

At the $\eta$-th iteration of FCS, the regression model is trained based on $\boldsymbol Z_{-j}^{(\eta)}[\mathcal{I}_{j,obs},:]$ and $\boldsymbol Z_j^{obs}$ by minimizing the loss, i.e.,
\begin{equation}
    \hat{\theta}_j^{(\eta)}=\arg \min\limits_{\theta_j}\sum_{i \in \mathcal{I}_{j,obs}} \ell( z_{i,j}-g_j(\boldsymbol z_{i,-j}^{(\eta)};\theta_j)),
\end{equation}
where $\ell(\cdot)$ is the mean squared error function.
Then, we predict a candidate value for each $\boldsymbol z_{i,-j}^{(\eta)}$ via the trained regression model, which is denoted as $\hat{z}_{i,j}^{(\eta)}$, i.e.,
\begin{equation}
    \hat{z}_{i,j}^{(\eta)}=g_j(\boldsymbol z_{i,-j}^{(\eta)};\hat{\theta}_j^{(\eta)}).
\end{equation}
Together, they are written in the form of a vector as $\hat{\boldsymbol Z}_j^{(\eta)}$, which is expressed as
\begin{equation}
    \hat{\boldsymbol Z}_j^{(\eta)}=[\hat{z}_{1,j}^{(\eta)},\hat{z}_{2,j}^{(\eta)},\cdots,\hat{z}_{N,j}^{(\eta)}]^\top.
\end{equation}
To impute $\boldsymbol Z_j^{mis}$, let us focus on each missing entry of it, for instance $z_{i_m,j},\ i_m \in \mathcal{I}_{j,mis}$, whose candidate is $ \hat{z}_{i_m,j}^{(\eta)}$.
Then we find $d$ nearest candidates from $\hat{\boldsymbol Z}_j^{(\eta)}[\mathcal{I}_{j,obs}]$ for which $|\hat{z}_{i_m,j}^{(\eta)}-\hat{ z}_{i,j}^{(\eta)}|, i_m \in \mathcal{I}_{j,mis}, i \in \mathcal{I}_{j,obs}$ is minimal.
Suppose the $d$ candidates are
\begin{align*}
    \hat{ z}_{i_1,j}^{(\eta)}, \hat{ z}_{i_2,j}^{(\eta)},\cdots,\hat{ z}_{i_d,j}^{(\eta)},\quad i_1, i_2,\cdots,i_d \in \mathcal{I}_{j,obs},
\end{align*}
which can be written in the form of  a set as $\mathcal{C}_{i,j}$, i.e., $\mathcal{C}_{i,j}=\{\hat{ z}_{i_1,j}^{(\eta)}, \hat{ z}_{i_2,j}^{(\eta)},\cdots,\hat{ z}_{i_d,j}^{(\eta)} \}$.
Finally, we obtain imputation for $z_{i,j},\ i \in \mathcal{I}_{j,mis}$ by sampling from $\mathcal{C}_{i_m,j}$ and denote it as $z_{i_m,j}^{(\eta)}$, i.e.,
\begin{equation}
    z_{i_m,j}^{(\eta)}\sim \mathcal{C}_{i,j},\quad i_m \in \mathcal{I}_{j,mis}.
\end{equation}
Indeed, the set $\mathcal{C}_{i_m,j}$ provides an empirical distribution for $z_{i_m,j},\ i_m \in \mathcal{I}_{j,mis}$. The operations described from (17) to (20) correspond the conceptual description in (12) and (13).
After all iterations, the final candidates corresponding to training dataset are denoted as $\hat{\boldsymbol Z}_j$, which are prepared for the use of sampling at the operational forecasting stage.
In particular, missing values can be directly imputed via (18) when only point forecasts are needed.

\subsection{Random forest}
Indeed, the model described in (16) can be specified as any regression model, such as linear regression, random forest, etc.
In this paper, it is specified as a random forest, as tree models usually perform well in practice \citep{januschowski2021forecasting}.
It grows $B$ regression trees, each of which is trained on bootstrap samples from training data.
Hence, the regression model that maps variables $\boldsymbol z_{-j}$ to $z_j$ is described as
\begin{equation}
    g_j(\boldsymbol z_{-j};\hat{\theta}_j)=\frac{1}{B}\sum_{b=1}^B g_{j,b}(\boldsymbol z_{-j}),
\end{equation}
where $g_{j,b}(\boldsymbol z_{-j})$ is a regression tree.
The splitting variable and splitting points of regression trees are often determined by the CART algorithm.
Details about the CART algorithm can be found in \citep{hastie01statisticallearning}.
Suppose we already have partitioned the variables into $M$ regions, i.e., $R_1,R_2,\cdots,R_M$.
And we model the target as a constant $c_m$ in each region.
The regression function is described as
\begin{equation}
    g_{j,b}(\boldsymbol z_{-j})=\sum_{m=1}^M c_m I(\boldsymbol z_{-j}\in R_m),
\end{equation}
where $I(\cdot)$ is the indicator function.
In particular, $c_m$ is estimated as the average of targets $z_j$ in the region $R_m$, i.e.,
\begin{equation}
    \hat{c}_m=\frac{1}{|\mathcal{I}_{R_m}|}\sum\limits_{i\in \mathcal{I}_{R_m}}z_{i,j},
\end{equation}
where $\mathcal{I}_{R_m}=\{i\ |\ \boldsymbol z_{i,-j}\in R_m\}$.
The model grows like a binary tree. To begin with, we consider the space is split at variable $Z_a,\ a\in \{1,\cdots,j-1,j+1,\cdots,pk+1 \}$ and point $s$, then we obtain two halves:
\begin{equation}
    R_1(a,s)=\{\boldsymbol z_{-j}|z_a\leq s \}, \ R_2(a,s)=\{\boldsymbol z_{-j}|z_a> s \}.
\end{equation}
It is fulfilled by a greedy algorithm, i.e.,
\begin{equation}
    \min \limits_{a,s}\left[\min\limits_{c_1}\sum \limits_{\boldsymbol z_{i,-j}\in R_1(a,s)}\ell(z_{i,j}-c_1)+\min\limits_{c_2}\sum \limits_{\boldsymbol z_{i,-j}\in R_2(a,s)}\ell(z_{i,j}-c_2)\right].
\end{equation}
Repeat the splitting process in the generated two regions, and stop only when minimum node size is reached.

\subsection{Forecasting Stage}

\begin{figure*}[!ht]
\centering
\includegraphics[width=6in]{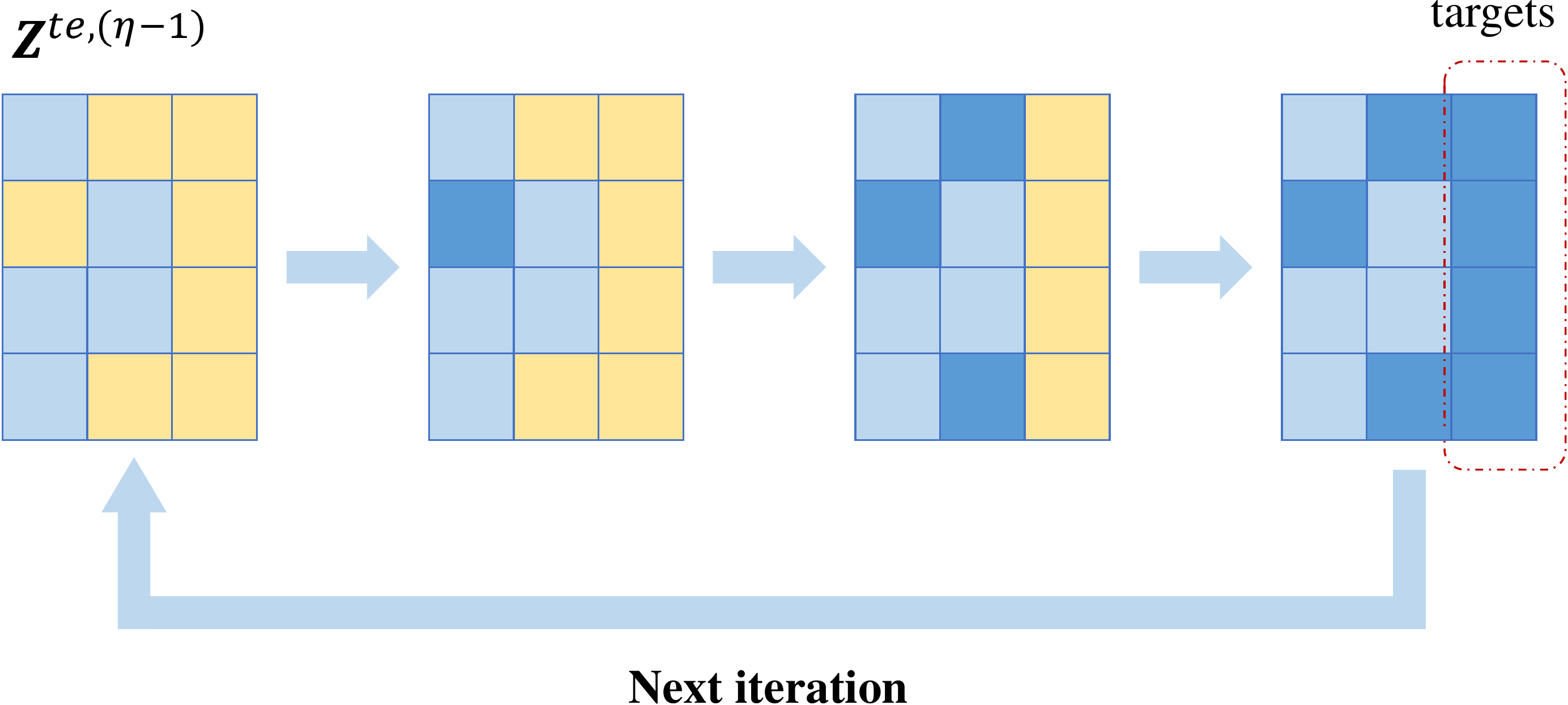}
\caption{Illustration of the $\eta$-th iteration at the operational forecasting stage. Light blue blocks indicate observations, yellow blocks indicate missing values, and dark blue blocks indicate imputation (forecasting).}
\label{forecast}
\end{figure*}

After training the imputation model, we obtain a collection of estimated random forests $\{g_j\}$ with parameters $\{\hat{\theta}_j\}$ and candidates $\{\hat{\boldsymbol Z}_j\}$.
At the operational forecasting stage, we feed sample $\boldsymbol z_t=[\boldsymbol x_t^{\top},y_t]^\top$ ($y_t$ is missing by default) into the estimated imputation model, and iteratively impute each missing value in $\boldsymbol z_t$ according to (11), (13)-(15), which is illustrated in Figure~\ref{forecast}.
Compared to the training stage, parameters are fixed now; thus we only conduct iterative imputation here.
Particularly, $L$ equally likely imputations for $\boldsymbol z_t$ are obtained, which are written as
\begin{align*}
    \Tilde{\boldsymbol z}_t^1,\Tilde{\boldsymbol z}_t^2,\cdots,\Tilde{\boldsymbol z}_t^L.
\end{align*}
Indeed, here $\boldsymbol z_t^{obs}=\boldsymbol x_t^{obs}$, $\boldsymbol z_t^{mis}=[{\boldsymbol x_t^{mis}}^\top, y_t]^\top$.
That is, $\boldsymbol z_t^{mis}$ is imputed by realizations from the estimated distribution $f_{X^{mis},Y|X^{obs}}(\boldsymbol z_t^{mis},y_t|\boldsymbol x_t^{obs};\mathcal{M},\hat{\Theta})$.
To get an empirical distribution for $f_{Y|X^{obs}}(y_t|\boldsymbol x_t^{obs};\mathcal{M},\hat{\Theta})$, we just fetch the corresponding value for $y_t$ in each $\Tilde{\boldsymbol z}_t^i$, i.e., the last entry of $\Tilde{\boldsymbol z}_t^i$, which is denoted as $\Tilde{y}_t^i$, i.e.,
\begin{equation}
    \Tilde{y}_t^i=\Tilde{z}_{t,pk+1}^{i},\quad i=1,\cdots,L.
\end{equation}
Recall that $y_t$ is the realization of the random variable $Y_{n,t+h}$, i.e., $\Tilde{y}_t^i$ is the realization from $f_{Y_{n,t+h}|t}(y|\boldsymbol x_t;\mathcal{M},\hat{\Theta})$.
Thus we rewrite $\Tilde{y}_t^i$ as $\Tilde{y}_{n,t+h|t}^i$, all of which form a set, i.e.,
\begin{align*}
    \{\Tilde{y}_{n,t+h|t}^1,\Tilde{y}_{n,t+h|t}^2,\cdots,\Tilde{y}_{n,t+h|t}^L \}.
\end{align*}
Besides, we note that (26) is a surrogate of (10), which serves as marginalization operation when $L$ is quite large.
The point forecast $\hat{y}_{n,t+h|t}$ is given as an average, which is expressed as
\begin{equation}
    \hat{y}_{n,t+h|t}=\frac{1}{L}\sum_{i=1}^L\Tilde{y}_{n,t+h|t}^i.
\end{equation}

\section{Simulation study}
Before validating the proposed approach on real data, we illustrate its applicability to point forecasting based on two related simulated processes, i.e., the autoregressive (AR) process and vector autoregressive (VAR) process.
The results are assessed in terms of root-mean-square error~(RMSE) here.
Let $\mathcal{I}_{y,obs}$ denote the indices of observations in the test set.
Then RMSE on the test set is described as
\begin{equation}
    {\rm RMSE}=\sqrt{\frac{1}{|\mathcal{I}_{y,obs}|}\sum_{t\in \mathcal{I}_{y,obs}}(y_{t}-\hat{y}_{t})^2},
\end{equation}
where $y_t$ denotes the observation at time $t$, $\hat{y}_t$ denotes the point forecast at time $t$, and $|\mathcal{I}_{y,obs}|$ is the number of observed samples in the test set.
In each case, we remove parts of generated data at random to simulate missingness, where the missing rate is varied from $5 \% $ to $50 \% $.
Situations where missing rates are larger than $50 \% $ are regarded impractical and thus not included in the study. 
Then, $80\%$ of data are split as the training set, whereas another $20\%$ of data are split as the test set for genuine forecasting validation.
The missingness simulation and model validation are replicated 100 times for each missing rate.

\subsection{AR process}

In this case, we model an AR process of order 2, i.e.,
\begin{align*}
    Y_t = \alpha_0 + \alpha_1 Y_{t-1} + \alpha_2 Y_{t-2} + \epsilon_t,
\end{align*}
where $\alpha_0$ is a constant, $\alpha_1$ and $\alpha_2$ are parameters, and $\epsilon_t$ is a white noise centered on 0.
Let us set $\boldsymbol \alpha^\top $ as $[1,\ 0.33,\ 0.5]^\top $, and $\epsilon_t$ to follow the Gaussian $\mathcal{N}(0,0.01)$.
Concretely, we simulate a time series of length 8760, corresponding to a year of data with 1-hour resolution, and present it in Figure~6 (a).
The input features $\boldsymbol x_t$ have 2 dimensions, and target $y_t$ has 1 dimension.
Specifically, the imputation model is trained by 10 iterations, as suggested by \citet{van2006fully}.
The RMSE values with respect to different missing rates are shown in Figure~6 (b).
Intuitively, missing values lead to an increase in RMSE, and higher missing rates lead to larger RMSE.

\begin{figure}[!ht]
\centering  
\subfigure[Simulated data]{   
\centering   
\includegraphics[width=3.1in,height=2.5in]{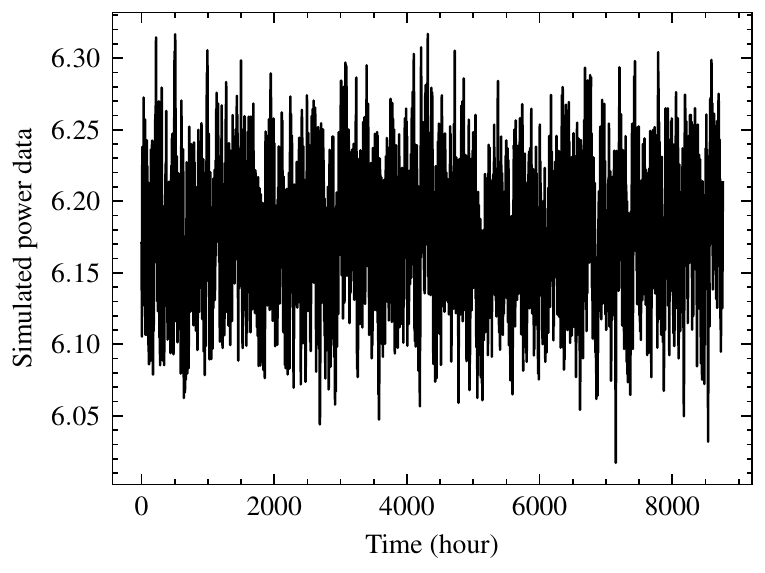}  
}
\subfigure[Results]{ 
\centering   
\includegraphics[width=3.1in,height=2.5in]{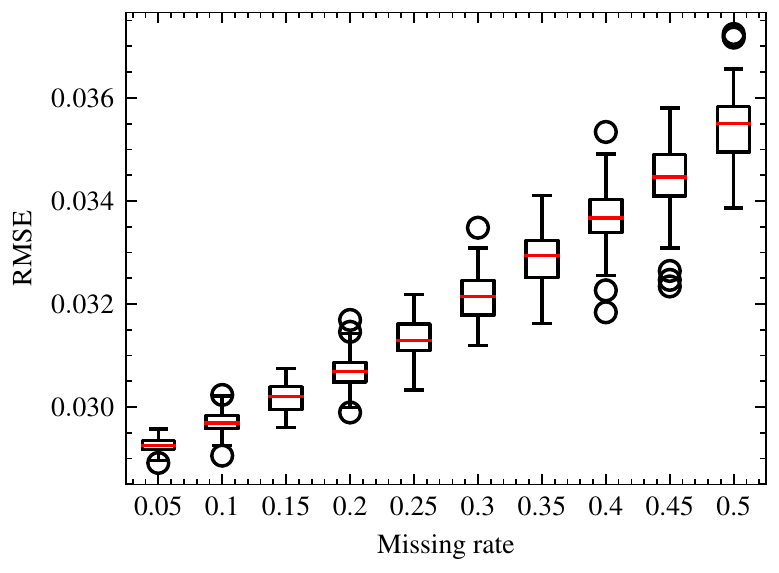}
}
\caption{(a) Example simulated time-series with an AR process (single replicate); (b) Box plot of 1-step ahead RMSE in the presence of missing values based on AR simulated data with respect to different missing rates (based on Monte-Carlo simulations with 100 replicates).}    
\label{AR}   
\end{figure}

Since experiments at each missing rate are replicated 100 times, we obtain the variance of RMSE at each missing rate.
As the missing rate increases, the variance of RMSE also increases, because the influence on training varies to a larger extent when the missing rate is high.

\begin{figure}[!ht]
\centering  
\subfigure[Simulated data]{   
\centering   
\includegraphics[width=0.47\textwidth]{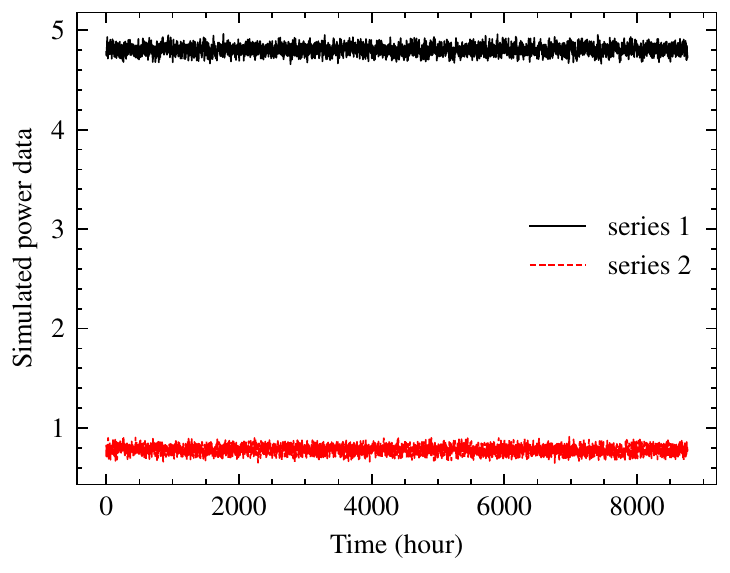} 
}
\subfigure[Features are from two series (no missing values in series 2)]{ 
\centering   
\includegraphics[width=0.47\textwidth]{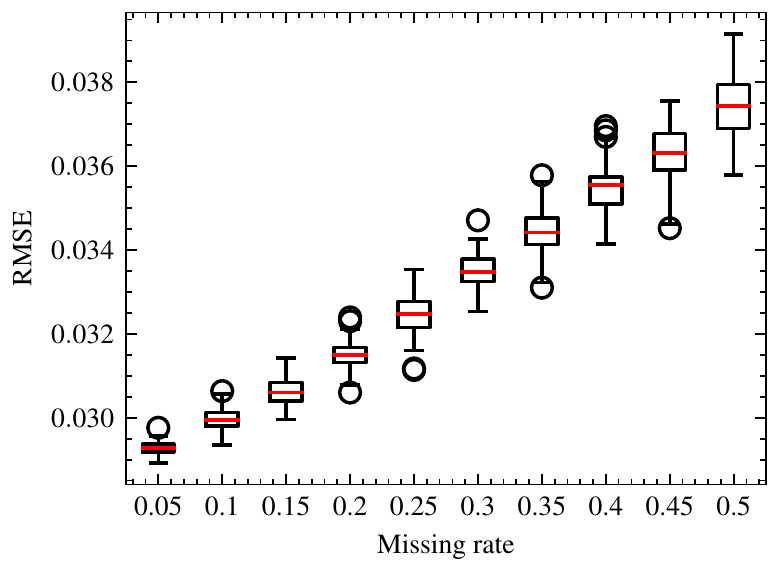}
}
\subfigure[Features are only from series 1]{ 
\centering   
\includegraphics[width=0.47\textwidth]{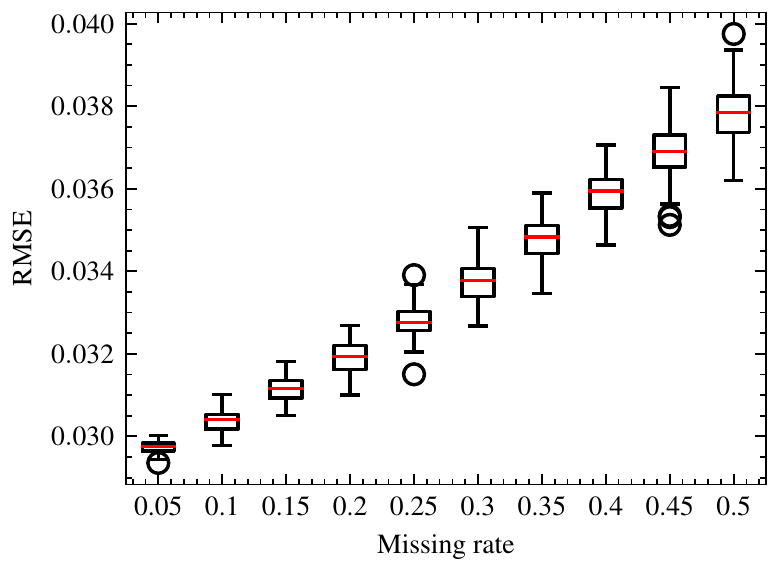}
}
\subfigure[Features are from two series (with missing values in series 2)]{ 
\centering   
\includegraphics[width=0.47\textwidth]{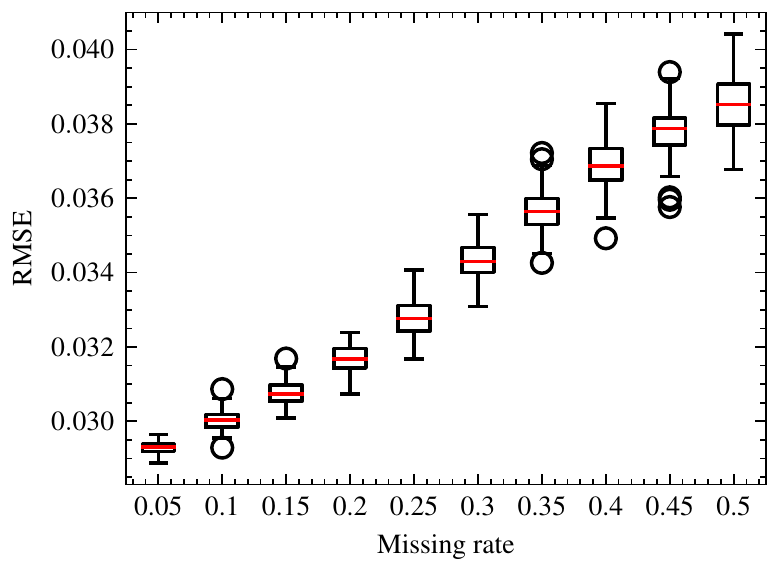}
}
\caption{(a) Simulated time series of the VAR process; (b) Box plot of 1-step ahead RMSE based on VAR simulated data with respect to different missing rates for series 1 (Monte-Carlo with 100 replications); (c) Box plot of 1-step ahead RMSE based on series 1 with respect to different missing rates for series 1 (Monte-Carlo with 100 replications); (d) Box plot of 1-step ahead RMSE based on VAR simulated data with respect to different missing rates for both series (Monte-Carlo with 100 replications).}    
\label{VAR}   
\end{figure}

\subsection{VAR process}

We model a VAR process of order 2, i.e., 
\begin{align*}
    Y_{1,t}=\alpha_{0}+\alpha_{1,1}Y_{1,t-1}+\alpha_{1,2}Y_{1,t-2}+\alpha_{2,1}Y_{2,t-1}+\alpha_{2,2}Y_{2,t-2}+\epsilon_{1,t},\\
    Y_{2,t}=\beta_{0}+\beta_{1,1}Y_{1,t-1}+\beta_{1,2}Y_{1,t-2}+\beta_{2,1}Y_{2,t-1}+\beta_{2,2}Y_{2,t-2}+\epsilon_{2,t},
\end{align*}
where $\boldsymbol \alpha_1^\top=[1, \ 0.88, \ -0.1, \ 0.15, \ -0.14]^\top$, $\boldsymbol \alpha_2^\top=[1, \ 0.69, \ -0.05, \ 0.07, \ -0.23]^\top$, $\epsilon_{1,t}\sim \mathcal{N}(0,0.01)$, and $\epsilon_{2,t}\sim \mathcal{N}(0,0.01)$.
We still simulate time series of length 8760 and present them in Figure~7(a). In this case, we focus on forecasting the future value of $Y_{1,t}$ by using previous realizations of both series. Now, the input features $\boldsymbol x_t$ have elements, whereas the target $y_t$ has a dimension of 1 only. We still train the imputation model with 10 iterations.

As a starting point, we assume there are no missing values in series $\{y_{2,t}\}$ and only vary the missing rate for $\{y_{1,t}\}$. The overall RMSE values are presented in Figure~7(b). As with the AR case, RMSE and its variance increase as the missing rate increases. For comparison, we consider two other scenarios, i.e., only using features from $\{y_{1,t}\}$, and simulating missingness for both $\{y_{1,t}\}$ and $\{y_{2,t}\}$, the results of which are respectively shown in Figure~7(c) and Figure~7(d).
Comparing Figure~7(b) and Figure~7(c), we observe that the RMSE in Figure~7(b) is lower, which translates into saying that forecasting can be improved by using information from correlated series. However, as shown in Figure~7 (d), the benefit of
using features from $\{y_{2,t}\}$ is still noticeable when the missing rate of $\{y_{2,t}\}$ is not too high.
When the missing rate of $\{y_{2,t}\}$ is higher than $30\%$, using features of $\{y_{2,t}\}$ will even hamper the performance.

\section{Case study}

Besides the above simulation study, we further validate our approach based on real-world data from the USA. The case study considers a typical forecasting setup, where some data is used for estimating model parameters (training set)  and the remainder of the data for genuine out-of-sample forecast verification (test set). Both point and probabilistic forecasting are considered. Also, since the dataset gathers data for multiple wind farms in a limited area, we can look at the case of employing univariate approaches (i.e., use of local data only), but also at a case where data from surrounding wind farms is used to improve forecasts. In that case, it is intuitively expected that one is further exposed to the likelihood and potential consequences of missing data. Note that the goal of this case study is not to pick and choose the best model and forecasting approach, but instead to show the impact of missing values on forecasting and the effectiveness of the proposed approach to accommodate those. In the following, we first describe the dataset and our experimental setup, the forecast verification framework and the benchmark approaches. The results obtained are then described and discussed. Codes and data\footnote{github link to be added in the final version %https://github.com/honglinwen/Forecasting-with-missing-values.git
} are publicly available.

\subsection{Data description}

Data from the USA are generated by the Wind Integration National Dataset (WIND) Toolkit \citep{draxl2015wind}, which are therefore not completely real but capture the dynamics of wind power generation.
Indeed, there are no missing values in this dataset.
Then, we randomly remove some values to simulate missingness, based on which all models are estimated and validated.
Concretely, the dataset contains 3 wind farms located in South Carolina, within a 150~km area.
The spatial-temporal dynamics among wind farms suggest that one could use data from nearby wind farms to improve the forecasts.
It gathers data over 7 years, from 2007 to 2013, with an hourly temporal resolution.
All wind power measurements are normalized by their corresponding capacities.
%Besides, we simulate different missingness for each wind farm.

\subsection{Experimental setup}

\subsubsection{Different types of case-studies}

Based on the data described above, we concentrate on both point and probabilistic forecasting in three different types of case studies, representing alternative approaches to forecasting (local data only, and with data from surrounding wind farms), as well as different types of missingness, i.e., sporadic and block missingness. We also consider forecasting with lead times from $k=1$ to 6 steps ahead. More precisely these cases can be described as:
\begin{description} 
    \item[Case 1:] Forecasting at a single site, using local data only (hence, with an autoregressive model). Data is missing sporadically and randomly, on a pointwise basis. The rates of missingness are $10\%$ and $20\%$, respectively, to investigate the performance of the approach conditional to how much data is missing.
    \item[Case 2:] Forecasting at a single site using local data only (hence, with an autoregressive model). Data is missing over given time intervals (block missingness) though at random. The number of blocks with missing data is set to 600. These are randomly located over the dataset. The length of the block with missing data is random and uniformly distributed between 5 to 30 time steps.
    \item[Case 3:] Forecasting at a chosen site, but using data from both that site and the nearby sites (hence, with a vector autoregressive model). The two types of missingness mentioned before (pointwise and block missingness) are considered.
\end{description}
In all 3 cases, when issuing a forecast at time $t$ for lead time $t+k$, lagged observations are used as input features (since using autoregressive models). As feature selection is not the focus of this paper, we performed a preliminary study to select lags based on training data. As a result, we work in the following with autoregressive models with the 6 lagged observations (so, from $t-5$ to $t$). The generalized logit-normal transform proposed by \citet{pinson2012very} is further employed as a pre-processing stage to accommodate the double-bounded nature of wind power generation time-series (i.e., nonlinear and with the variance of residuals conditional upon the mean level).

\subsubsection{Forecast verification: relevant scores and diagnostic tools}

The quality of point forecasts is commonly evaluated with an RMSE criterion (consistent with the use of a quadratic loss in learning and forecast verification), whereas the quality of probabilistic forecasts is most often assessed by using the Continuous Ranked Probability Score~(CRPS). Given a lead time $h$, we denote the cumulative density function for wind power generation $ Y_{t+h} $, predicted at time $t$ for time $t+h$, as $ F_{t+h} $. Then, the CRPS for the predicted $ F_{t+h}$ and corresponding observation $y_{t+h}$ is defined as
\begin{equation}
    {\rm CRPS} (F_{t+h},y_{t+h})=\int_y \big( F_{t+h}( y)-\mathbbm{1}(y-y_{t+h}) \big)^2 dy,
\end{equation}
where $\mathbbm{1}(\cdot)$ is a unit step function at the location of the observation $y_{t+h}$ (also known as a Heaviside function), which can be regarded as the empirical cumulative density function of the observation $y_{t+h}$. Eventually, given the lead time $h$, we report the average CRPS value over all forecast-verification pairs, i.e.,
\begin{equation}
    {\rm CRPS}_h=\frac{1}{|\mathcal{I}_{y,obs}|}\sum_{t+h\in \mathcal{I}_{y,obs}}{\rm CRPS}(F_{t+h},y_{t+h}).
\end{equation}

Besides the use of a proper skill score like the CRPS, informing about the overall skill and quality of the probabilistic forecasts (in the form of predictive densities), we will assess the probabilistic calibration of the predictive densities with reliability diagrams. For an extensive description of such reliability diagrams and their use in the assessment of probabilistic calibration, the reader is referred to \citet{pinson2010reliability}. In parallel, in order to see how the probabilistic forecasts concentrate information, their sharpness is evaluated by calculating the width of central prediction intervals. I.e., for a given nominal coverage rate $1-\beta$, these central prediction intervals are bounded by quantiles with nominal levels $\beta/2$ and $1-\beta/2$. For a general overview of probabilistic forecast verification, see \citet{gneiting2007}.

\subsubsection{Benchmarks}

In general, we use three categories of benchmarks, i.e., the climatology/persistence method, an ITP approach, and a UI approach with a distributional assumption.
For point forecasting, persistence uses the latest observation as the forecast. To implement the ITP approach, we respectively use mean imputation and advanced regression-based imputation namely MissForest \citep{stekhoven2012missforest} in the pre-processing procedure and employ a random forest as the backbone regression model, which are abbreviated as RF-M and RF-R respectively.
And, the state-of-the-art model DeepAR \citep{salinas2020deepar} is adopted, which uses intermediate results of the long-short term memory model to impute missing values at both model estimation and operational stages.
The copula-based imputation model proposed by \citet{zhao2020missing} is adopted to implement the UI approach.
It is also a multiple imputation model, though relying on a distributional assumption. The retraining approach\citep{tawn2020missing} is used as a benchmark model, which consists in retraining the model without missing features. Besides, we consider a reference model that is implemented by a random forest and trained based on the complete dataset, which is abbreviated as RF-C. The benchmark models for point forecasting, as well as used abbreviations, are gathered in Table~\ref{tab:Tableabbrv}.

\begin{table}[!ht]
\caption{Abbreviations for point and probabilistic forecasting benchmark models.}
\small
 \label{tab:Tableabbrv}
 \centering
\begin{threeparttable}
 \begin{tabular}{ll}
 \hline
 Abbreviation & Description (point forecasting) \\ 
 \hline
 RF-M & Random forest with the mean imputation as preprocessing  \\ 
 RF-R & Random forest with the regression-based imputation as preprocessing\\ 
 Copula & Copula-based imputation model within universal imputation strategy \\
 DeepAR & Deep learning model that uses intermediate results to impute missing values \\
 RF-C & Random forest trained based on the complete dataset \\
 \hline
 Abbreviation & Description (probabilistic forecasting) \\ 
 \hline
 Gauss-M & Gaussian model with the mean imputation as preprocessing  \\ 
 Gauss-R & Gaussian model with the regression-based imputation as preprocessing\\ 
 Copula & Copula-based imputation model within universal imputation strategy \\
 DeepAR & Deep learning model that uses intermediate results to impute missing values \\
 QR-R & QR model with the regression-based imputation as preprocessing \\
 QR-C & QR model trained based on the complete dataset \\
\hline
  \end{tabular}
\end{threeparttable}
\end{table}

As for probabilistic forecasting, climatology is set as a naive benchmark. It utilizes the empirical distribution of all historical values to communicate the probability distribution of future wind power generation. To implement the ITP approach, a model with the Gaussian distributional assumption as well as a QR model are adopted as backbone models.
In particular, the base model chosen for QR is the gradient boosting machine, which supports QR and ranks highly on leaderboards of recent forecast competitions \citep{januschowski2021forecasting}, including the GEFCom 2014 \citep{LANDRY20161061} for instance. For the model with the Gaussian distributional assumption, we use a neural network to estimate the shape parameters of Gaussian distributions. The QR model with regression-based imputation as preprocessing is abbreviated as QR-R, while the Gaussian models with mean and regression-based imputation are abbreviated as Gauss-M and Gauss-R. Again, the DeepAR model \citep{salinas2020deepar} is used as a benchmark, since it is allowed to communicate Gaussian densities.
The UI approach is still implemented via the copula-based model. Besides, we set the QR model trained based on the complete dataset as a reference, which is abbreviated as QR-C. The benchmark models for probabilistic forecasting, as well as corresponding abbreviations, are also collated in Table~\ref{tab:Tableabbrv}.

\subsection{Results and discussion}
Results that correspond to the aforementioned three cases are respectively reported in three different subsections and followed by further discussion.

\subsubsection{Case 1}

Emphasis is first placed on sporadic missingness, i.e., for the case where single values are missing, at random times. Let us start by presenting and discussing results for the most severe rate of missingness, of $20\%$. In practice, this means that 20\% of the values are missing, at random locations over both training and testing sets. The point forecasting results in terms of RMSE are collated in Table~\ref{tab:Table 1}.

\begin{table}[!ht]
\caption{RMSE values as a function of the lead time (Case 1, missing rate of 20\%). RMSE values are expressed in percentage of normalized capacity.}
\small
 \label{tab:Table 1}
 \centering
\begin{threeparttable}
 \begin{tabular}{cccccccc}
 \hline
 Lead Time (steps) & Persistence & RF-M  & RF-R & Copula & FCS & DeepAR & RF-C\\ 
 \hline

 1 & 16.8 & 17.7 & 16.1 & 17.3 & 15.9 & 17.0 & 14.6 \\ 
 2 & 21.1 & 20.9 & 19.8 & 21.3 & 19.5 & 20.6 & 18.9\\ 
 3 & 24.7 & 23.4 & 22.5 & 24.3 & 22.3 & 23.3 & 21.9\\ 
 6 & 32.7 & 28.2 & 28.0 & 30.6 & 27.9 & 29.3 & 27.6\\ 

 \hline
 \end{tabular}
\end{threeparttable}
\end{table}

Not surprisingly, the RMSE increases with the lead time, and, forecast quality in the presence of missing values is worse than when there is no missing data. There, missing values have a negative impact at both model estimation and operational forecasting stages. Persistence is a competitive benchmark, as it is easy to implement and the resulting forecast quality is difficult to outperform for such short lead times. In parallel, RF-M performs worse than persistence for 1-step ahead forecasts, most likely due to errors in imputation introduced by this pre-processing procedure. Given training datasets $\boldsymbol X^{tr}$ and $\boldsymbol Y^{tr}$, one can estimate a regression model $f^P$ that is equivalent to the reference model, if the imputed datasets are as same as the real complete datasets $\boldsymbol X^{tr,P}$ and $\boldsymbol Y^{tr,P}$. However, the imputed datasets usually deviate from the real complete datasets. Then, the model estimated based on $\boldsymbol X^{tr,C}$ and $\boldsymbol Y^{tr,C}$, denoted as $f^C$, is different from $f^P$. That is, the closer the imputed datasets are to the real complete datasets, the closer $f^C$ is to $f^P$.

Obviously, RF-R has better performance than RF-M, since the regression-based imputation is superior to the mean imputation. Besides, at the operational forecasting stage, it is still required to impute input features, which may also accumulate errors. Although DeepAR is free of any pre-processing stage, the imputed values may still deviate from real values, introducing errors to both the model training and forecasting stages. As shown in Table~\ref{tab:Table 1}, the performance of DeepAR is even worse than the simple benchmark model, i.e., RF-R.

The used copula- and FCS- based models fall into the category of UI approach.
Compared to the ITP approach, the UI approach has the advantage that it is free of a pre-processing procedure, which avoids introducing errors aroused by the pre-processing procedure into the forecasting task.
The FCS-based model outperforms RF-M and RF-R, while the copula-based model is inferior to them, as revealed in Table~\ref{tab:Table 1}.
Although the copula method allows to characterize several kinds of distributions, it is required to specify the transform function here, which means that a specific distributional assumption is implied.
This may impede the performance of the copula-based model when the distributional assumption cannot fit the underlying distribution well.
By contrast, the FCS method is free of such an assumption and therefore has a better performance than the copula-based model.
It suggests that by using a distribution-free imputation method like FCS, the UI approach is superior to the ITP approach.

Besides, we compare our proposed approach and the retraining approach discussed by \citet{tawn2020missing} by focusing on a specific missing pattern (i.e., the last feature is missing), and present the RMSE values in Table~\ref{tab:Table 2}. It is intuitive that the performance of the retraining approach is comparable to RF-M and FCS, as the retraining approach is free of a pre-processing stage and utilizes a complete set of observations to estimate the parameters. Eventually, the quality of the forecasts is linked to the informative value of the features retained. However, it also implies that a specific model is needed for each missingness pattern. Then, a specific training dataset is required for each pattern, which means only parts of the data are used to estimate a model. Besides, the retraining approach will suffer the curse of dimensionality. That is, denoting the dimension of features as $d$, the retraining approach will independently train $2^d$ models. While the training time for a set of point forecasting models may be acceptable, the computational costs will steeply increase for probabilistic forecasting cases. In contrast, the proposed UI approach is not only free of any pre-processing stage but also applicable to all missingness patterns once trained.

\begin{table}[!ht]
\caption{RMSE for 1-step ahead forecasts (Case 1, last feature missing). RMSE values are expressed in percentage of normalized capacity.}
\small
 \label{tab:Table 2}
 \centering
\begin{threeparttable}
 \begin{tabular}{cccccccc}
 \hline
 Lead Time (steps) & Persistence & RF-M  & RF-R & Copula & FCS & DeepAR & Retraining \\ 
 \hline

 1 & 15.8 & 16.2 & 15.1 & 15.9 & 15.0 & 15.8 & 15.3 \\ 

 \hline
 \end{tabular}
\end{threeparttable}
\end{table}

Next, we move on to the results for probabilistic forecasting, with the CRPS values obtained collated in Table~\ref{tab:Table 3}. Here, missing values have nearly no influence on the performance of climatology, since climatology characterizes uncertainty based on the empirical distribution of all historical observations. This distribution is not highly modified when a fairly limited number of samples are missing.

\begin{table}[!ht]
\caption{CRPS  as a function of the lead time (Case 1, missing rate of 20\%). CRPS values are expressed in percentage of normalized capacity.}
\small
 \label{tab:Table 3}
 \centering
\begin{threeparttable}
 \begin{tabular}{ccccccccc}
 \hline
  {\scriptsize Lead Time (steps)} & Climatology & Gauss-M  & Gauss-R & QR-R & Copula & FCS & DeepAR& QR-C\\ 
 \hline

  1 & 18.6 & 9.2 & 7.5 & 7.8 & 11.5 & 6.9 & 7.8& 6.9 \\ 
  
  2 & 18.6 & 11.2 & 9.9 & 9.9 & 14.6 & 9.1 & 10.2& 9.3 \\
  
  3 & 18.6 & 12.7 & 11.7 & 11.7 & 17.0 & 10.9 & 12.1& 11.2\\
  
  6 & 18.6 & 15.9 & 15.5 & 15.4 & 22.4 & 14.7 & 16.5 & 15.1 \\

 \hline
 \end{tabular}
 
\end{threeparttable}
\end{table}

Comparing the Gauss-M and Gauss-R, we know that a better imputation method is still preferred by the ITP strategy in the context of probabilistic forecasting.
Both the Gauss-R and QR-R use the regression-based imputation as preprocessing procedure.
But they differ in backbone models -- Gauss-R relies on the Gaussian distributional assumption, whereas QR-R is distribution-free. 
Their performance is comparable in this case, which is different from the usual situation (i.e., complete datasets) where QR is always superior.
Obviously, one needs to estimate the shape parameters of Gaussian distribution in Gauss-R, but the parameters of several quantile functions in QR-R.
The parallel estimation of QR-R models may result in more errors in the ultimate estimated distribution.
Therefore, results are governed by both models and the influence of missing values on model estimation.
The performance of DeepAR is slightly worse than that of Gauss-R and QR-R, which suggests handling missing values in forecasting is nontrivial. Values imputed by the intermediate results of the model may also introduce errors at the model estimation stage.
The FCS-based model outperforms Gauss-R and QR-R, whereas the performance of the copula-based model is worse than those of Gauss-R and QR-R, which suggests that the distributional assumption may impede the performance of the UI approach.
Besides, the performance of the FCS-based model is comparable to that of the reference QR model trained based on the complete dataset, which validates the effectiveness of the FCS-based model.

\begin{figure}[!ht]
\subfigure[FCS-based model]{\includegraphics[width=0.48\textwidth]{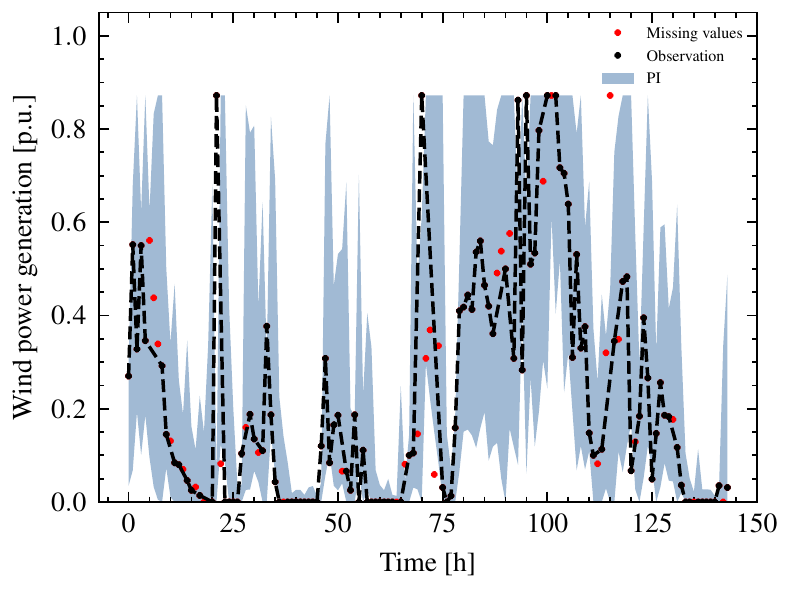}}
\subfigure[Reference model]{\includegraphics[width=0.48\textwidth]{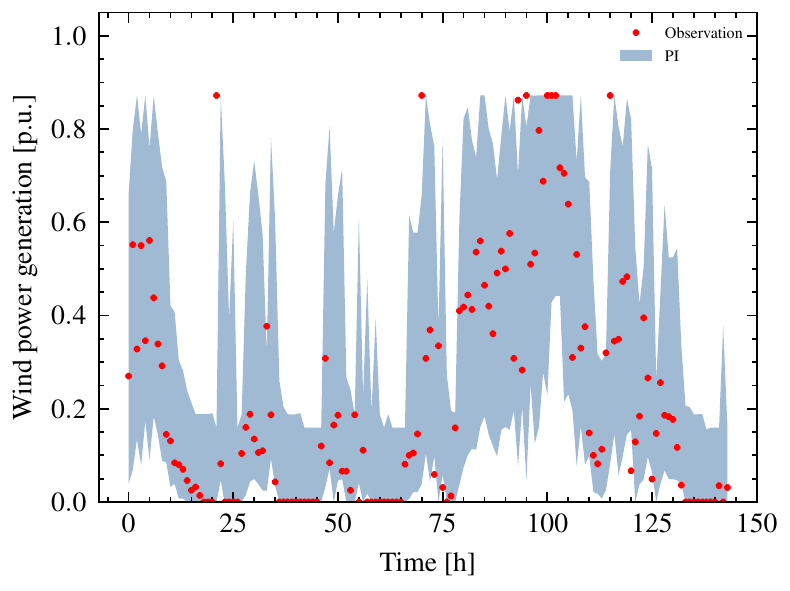}}
\caption{Illustration of 1-step ahead $90\%$ central prediction intervals over a period of 6 days, as issued by the FCS-based model (a), and the reference model (b). \label{PIexample}}
\end{figure}

We present the $90\%$ PIs of 6 days issued by the FCS-based and reference models, respectively, in Figure~\ref{PIexample}.

\begin{figure}[!ht]
\centering
\subfigure[Reliability diagrams]{\includegraphics[width=0.48\textwidth]{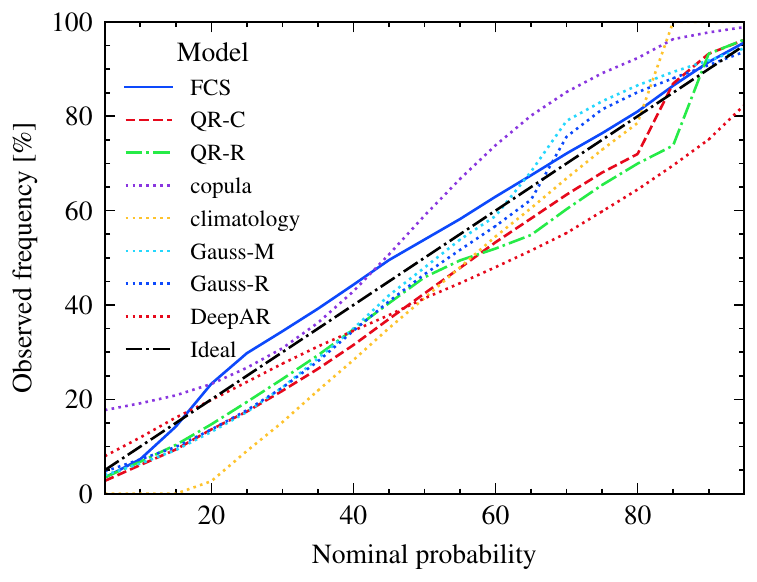}}
\subfigure[Sharpness diagrams]{\includegraphics[width=0.48\textwidth]{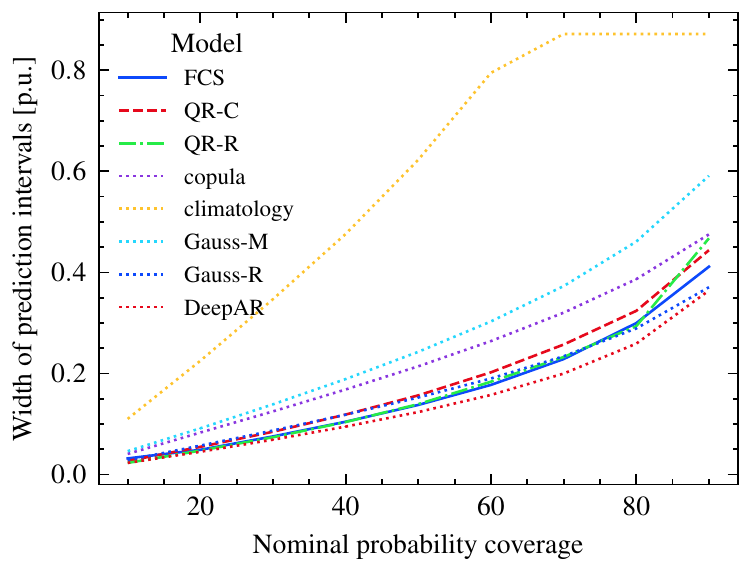}}
\caption{Assessment of 1-step ahead probabilistic forecasts for all models for Case 1, based on reliability diagrams (a) and sharpness diagrams (b).}
\label{probforecase1}
\end{figure}

Although the FCS-based approach encounters missing values at both model estimation and operational forecasting stages, its PIs are similar to those of the reference model. Specifically, at some periods e.g. from 35-h to 45-h, the prediction interval issued by the FCS-based approach are actually sharper than those of the reference model. A reliability assessment through the use of reliability diagrams is given in Figure~\ref{probforecase1}(a), while a sharpness assessment is performed by looking at the width of central prediction intervals (as a function of their nominal coverage rate), and depicted in Figure~\ref{probforecase1}(b). Models based on an ITP strategy tend to underestimate lower quantiles, while the reliability of DeepAR and copula-based model deviates from the ideal case to some extent. The FCS-based model achieves a level of  reliability and sharpness that is comparable to the reference model. The average of absolute values of deviations from perfect reliability is shown in Table~\ref{tab:Table reliability 1}. The deviation of the FCS is even smaller than that of the reference model, which is likely due to that the FCS is robust to overfitting.

\begin{table}[!ht]
\caption{The average of absolute values of deviations from perfect reliability in Case 1 (in percent).}
\small
 \label{tab:Table reliability 1}
 \centering
\begin{threeparttable}
 \begin{tabular}{ccccccccc}
 \hline
  Lead Time (steps) & Climatology & Gauss-M  & Gauss-R & QR-R & Copula & FCS & DeepAR& QR-C\\ 
 \hline

  1 & 9.20 & 3.37 & 3.52 & 4.81 & 4.81 & 2.78 & 10.57& 6.04 \\

 \hline
 \end{tabular}
 
\end{threeparttable}
\end{table}

The RMSE and CRPS values, when considering a missing rate of $10\%$, are collated in Table~\ref{tab:Table 4} and Table~\ref{tab:Table 5}, respectively. Compared to the results with a missing rate of $20\%$, the quality of the forecasts is improved. For point forecasting, the performance of RF-R is comparable to that of FCS, which means that the ITP strategy may be more acceptable when the missing rate is not that high. This is while, in the context of probabilistic forecasting, FCS still outperforms other models, which suggests that missing values may pose greater challenges to probabilistic forecasting. Besides, the performance of FCS is even better than that of QR-C, possibly hinting at the fact that FCS is less prone to overfitting.

\begin{table}[!ht]
\caption{RMSE values with different lead times in Case 1 on the condition that missing rate is $10\%$ (percentage of normalized capacity).}
\small
 \label{tab:Table 4}
 \centering
\begin{threeparttable}
 \begin{tabular}{cccccccc}
 \hline
 Lead Time (steps) & Persistence & RF-M & RF-R & Copula & FCS & DeepAR & RF-C\\ 
 \hline

 1 & 15.9 & 15.9 & 15.2 & 16.6 & 15.1 & 16.2 & 14.6 \\ 
 2 & 20.6 & 19.7 & 19.2 & 20.8 & 19.1 & 20.2 & 18.9\\ 
 3 & 24.4 & 22.5 & 22.1 & 24.1 & 22.0 & 23.3 & 21.9\\ 
 6 & 32.5 & 27.9 & 27.8 & 30.5 & 27.8 & 29.5 & 27.6\\ 

 \hline
 \end{tabular}
\end{threeparttable}
\end{table}

\begin{table}[!ht]
\caption{CRPS values with different lead times in Case 1 on the condition that missing rate is $10\%$ (percentage of normalized capacity).}
\small
 \label{tab:Table 5}
 \centering
\begin{threeparttable}
 \begin{tabular}{ccccccccc}
 \hline
  { \scriptsize Lead Time (steps)} & Climatology & Gauss-M  & Gauss-R & QR-R & Copula & FCS & DeepAR& QR-C\\ 
 \hline

  1 & 18.6 & 8.1 & 7.2 & 7.4 & 11.2 & 6.6 & 7.4& 6.9 \\ 
  
  2 & 18.6 & 10.5 & 9.6 & 9.6 & 14.3 & 8.9 & 9.9& 9.3 \\
  
  3 & 18.6 & 12.1 & 11.5 & 11.5 & 16.9 & 11.9 & 9.7& 11.2 \\
  
  6 & 18.6 & 15.9 & 15.3 & 15.3 & 22.4 & 14.7 & 16.6 & 15.1 \\

 \hline
 \end{tabular}
\end{threeparttable}
\end{table}

\subsubsection{Case 2}

In contrast to the sporadic missingness of Case~1, we simulate here missing values that span over time intervals (hence, referred to as block missingness). Remember that 600 blocks are randomly spread over the whole dataset, with lengths between 5 and 30 time steps. Let us first analyze and discuss point forecasting results. As a basis, the RMSE values of the points forecasts obtained with the different approaches are gathered in Table~\ref{tab:Table 6}.

\begin{table}[!ht]
\caption{RMSE values with different lead times in Case 2 (percentage of normalized capacity).}
\small
 \label{tab:Table 6}
 \centering
\begin{threeparttable}
 \begin{tabular}{cccccccc}
 \hline
 Lead Time (steps) & Persistence & RF-M  & RF-R & Copula & FCS & DeepAR & RF-C\\ 
 \hline

 1 & 15.8 & 14.9 & 14.8 & 16.4 & 14.9 & 15.9 & 14.6 \\ 
 2 & 20.9 & 19.2 & 19.2 & 21.1 & 19.3 & 21.0 & 18.9\\ 
 3 & 24.8 & 22.2 & 22.2 & 24.3 & 22.4 & 24.1 & 21.9\\ 
 6 & 32.9 & 27.9 & 27.9 & 30.6 & 28.1 & 30.9 & 27.6\\ 

 \hline
 \end{tabular}
\end{threeparttable}
\end{table}

The performance of RF-M is comparable to that of RF-R, most likely due to the fact that most samples here are complete.
In contrast, the copula-based model performs much worse than both RF-M and RF-R. Indeed, the estimation stage for the copula-based model is based on an expectation-maximization algorithm, which is sensitive to samples whose values are entirely missing. This may suggest that the copula-based model is not applicable to the situation of block missingness. Certainly, the samples whose values are entirely missing contain no information and can be deleted at the model estimation stage. As with Case~1, the performance of DeepAR is worse than that of RF-R, which reveals a caveat for the existing DeepAR framework in handling missing values. The performance of the FCS-based approach is comparable to that of RF-M/RF-R. One may then infer that ITP and UI types of strategies perform fairly similarly for point forecasting when experiencing block missingness. However, the picture will look different when extending the study to probabilistic forecasting. To assess the performance of the various approaches for that probabilistic forecasting case, we first look at CRPS values, which are gathered in Table~\ref{tab:Table 7}.

\begin{table}[!ht]
\caption{CRPS values with different lead times in Case 2 (percentage of normalized capacity).}
\small
 \label{tab:Table 7}
 \centering
\begin{threeparttable}
 \begin{tabular}{ccccccccc}
 \hline
  { \scriptsize Lead Time (steps)} & Climatology & Gauss-M  & Gauss-R & QR-R & Copula & FCS & DeepAR& QR-C\\ 
 \hline
  1 & 18.6 & 7.0 & 6.9 & 7.2 & 11.2 & 6.5 & 7.1& 6.9 \\ 
  2 & 18.6 & 9.8 & 9.6 & 9.7 & 14.5 & 9.0 & 10.2& 9.3 \\
  3 & 18.6 & 11.8 & 11.8 & 11.6 & 17.1 & 10.9 & 12.6& 11.2 \\
  6 & 18.6 & 15.7 & 15.7 & 15.6 & 22.4 & 14.9 & 17.9 & 15.1 \\
 \hline
 \end{tabular}
\end{threeparttable}
\end{table}

Not surprisingly, the performance of Gauss-R is slightly superior to that of Gauss-M. But their difference is smaller than what was observed in Case~1. It could be inferred that it is difficult to handle block missingness via imputation techniques. In the context of block missingness, regression-based imputation will also tend to impute missing values with mean values. Still, the performance of Gauss-R is comparable to that of QR-R. The FCS-based model yields the best performance. Combined with the results of Case 1, it indicates that this approach seems to be superior for different types of missingness, here both sporadic and block missingness. A clear point is that ITP strategies are highly sensitive to samples whose values are entirely missing since the rationale of ITP strategies is to utilize observed parts of samples to infer the missing parts. If a sample is completely unobserved, no information could be used for learning and eventually forecasting.

\begin{figure}[!ht]
\centering
\subfigure[Reliability diagrams]{\includegraphics[width=0.48\textwidth]{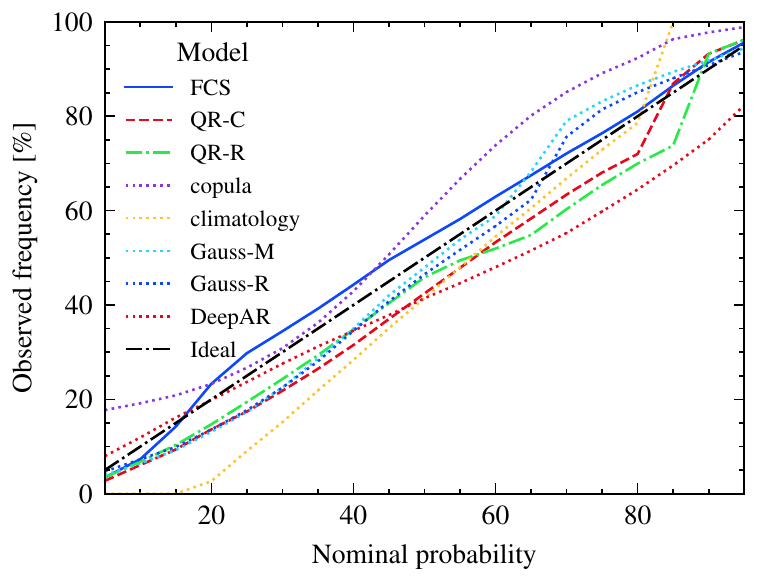}}
\subfigure[Sharpness diagrams]{\includegraphics[width=0.48\textwidth]{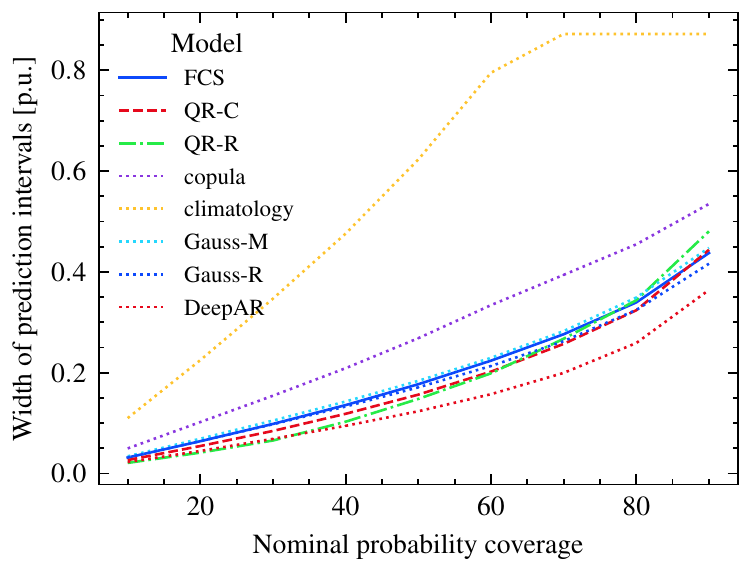}}
\caption{Assessment of 1-step ahead probabilistic forecasts for all models for Case 2, based on reliability diagrams (a) and sharpness diagrams (b).}
\label{probforecase2}
\end{figure}

Both reliability and sharpness are evaluated in Figure~\ref{probforecase2}, for 1-step ahead probabilistic forecasts (with reliability diagrams in Figure~\ref{probforecase2}(a) and sharpness diagrams in Figure~\ref{probforecase2}(b). The FCS-based approach achieves acceptable probabilistic calibration, especially in the case of lower and higher quantiles. As summary statistics, the average deviation (in absolute value) for perfect reliability is given in Table~\ref{tab:Table reliability 2}, for all approaches. There again, one verifies that the FCS-based approach yields the lowest deviation. In parallel, the prediction interval width for QR, Gaussian-based and  FCS-based approaches are very close, for all nominal coverage rates. The prediction interval width for DeepAR is somewhat smaller, though at the price of poorer probabilistic calibration. This is also reflected by the larger CRPS values for DeepAR, compared to the FCS-based approach.

\begin{table}[!ht]
\caption{Average deviation (in absolute value) from perfect reliability in Case 2 (in percent).}
\small
 \label{tab:Table reliability 2}
 \centering
\begin{threeparttable}
 \begin{tabular}{ccccccccc}
 \hline
  { \scriptsize Lead Time (steps)} & Climatology & Gauss-M  & Gauss-R & QR-R & Copula & FCS & DeepAR& QR-C\\ 
 \hline

  1 & 9.21 & 4.47 & 4.27 & 6.01 & 8.29 & 2.69 & 8.37 & 6.04 \\

 \hline
 \end{tabular}
 
\end{threeparttable}
\end{table}

\subsubsection{Case 3}

\begin{figure}[!ht]
\centering
\subfigure[Sporadic missingness case]{\includegraphics[width=0.48\textwidth]{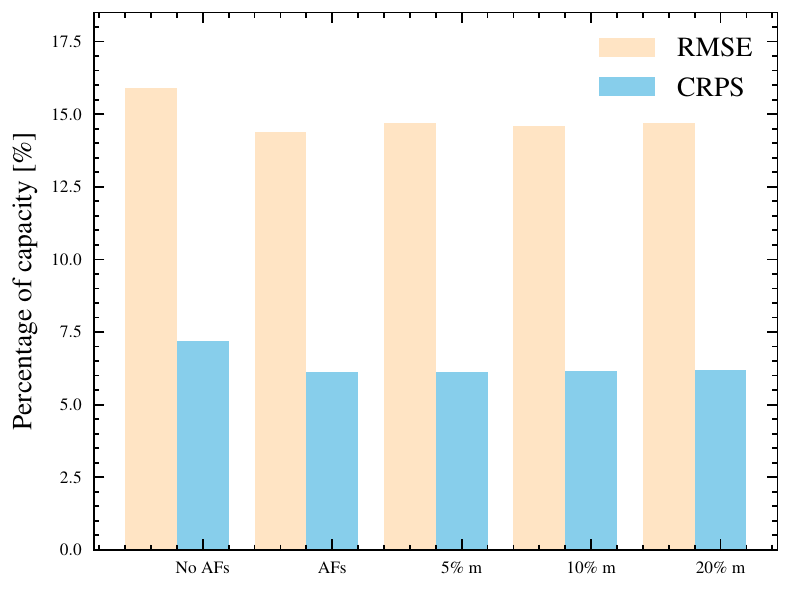}}
\subfigure[Block missingness case]{\includegraphics[width=0.48\textwidth]{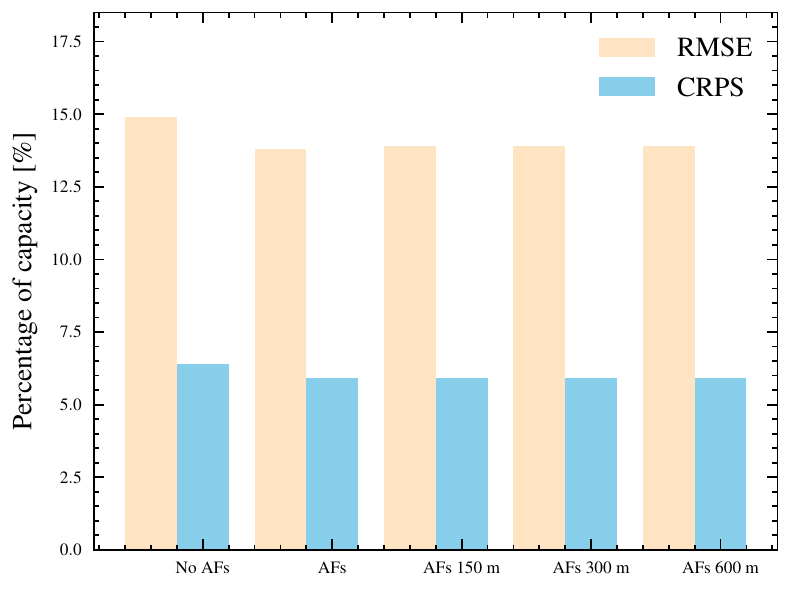}}
\caption{Assessment of 1-step ahead probabilistic forecasts for Case 3, in context of sporadic missingness (a) and block missingness (b).}
\label{probforecase3}
\end{figure}

In this subsection, we show that forecasting in the presence of missing values can still be improved by utilizing information on nearby sites as auxiliary features (AFs).
Besides input features of the chosen wind farm, we use previous wind power generation values from two nearby wind farms as AFs.
It is assumed that the missingness of nearby wind farms is different from the target wind farm, which is practical since missingness is usually caused by sensor faults or communication errors.
We consider both sporadic missingness and block missingness here. Particularly, we concentrate on 1-step ahead forecasts and investigate the impacts of different missing rates or missing blocks in AFs. The RMSE values in the context of sporadic missingness are shown in Figure~\ref{probforecase3}(a) (`AFs $p\%$ m' means $p\%$ of auxiliary features are missing), where we simulate different missing rates at two nearby wind farms and set the missing rate at the target wind farm as $20\%$.

As expected, the accuracy of point forecasting is improved with the assistance of AFs, which is comparable to RF-C in Table~1.
Furthermore, it can be seen that the benefit of AFs is robust since the performance is relatively consistent as the missing rate of AFs increases. It might be explained by the fact that the key information for forecasting comes from the target wind farm itself. So, it may not make a big difference when a few auxiliary features are missing.
The results of probabilistic forecasting are also shown in Figure~\ref{probforecase3}(a), which also suggests that AFs provide extra information and thus contribute to improving probabilistic forecasts. The RMSE and CRPS values in the context of block missingness are presented in Figure~\ref{probforecase3}(b) (`AFs $c$ m' means there are $c$ missing blocks in auxiliary features), where we simulate 600 missing blocks at the target wind farm. It is seen that auxiliary features can still improve the quality of the forecasts in that case.

\begin{comment}

\begin{table}[!ht]
\caption{RMSE and CRPS values in Case 3 and in the context of sporadic missingness (percentage of normalized capacity).}
\small
 \label{tab:Table 8}
 \centering
\begin{threeparttable}
 \begin{tabular}{cccccc}
 \hline
 & No AFs  & AFs & AFs $5\%$ m & AFs $10\%$ m & AFs $20\%$ m\\ 
 \hline

 RMSE & 15.9 & 14.4 & 14.7 & 14.6 & 14.7 \\ 
 CRPS & 6.9 & 6.1 & 6.1 & 6.2 & 6.2\\ 

 \hline
 \end{tabular}

\end{threeparttable}
\end{table}

\begin{table}[!ht]
\caption{RMSE and CRPS values in Case 3 and in the context of block missingness (percentage of normalized capacity).}
\small
 \label{tab:Table 9}
 \centering
\begin{threeparttable}
 \begin{tabular}{cccccc}
 \hline
 & No AFs  & AFs & AFs 150 m & AFs 300 m & AFs 600 m\\ 
 \hline

 RMSE & 14.9 & 13.8 & 13.9 & 13.9 & 13.9 \\ 
 CRPS & 6.4 & 5.9 & 5.9 & 5.9 & 6.0\\ 

 \hline
 \end{tabular}
\end{threeparttable}
\end{table}
\end{comment}

\subsubsection{Training time}

We note that by using the FCS method, the proposed UI approach is always superior in the context of probabilistic forecasting.
However, it costs much time to perform Gibbs sampling to provide probabilistic forecasting.
The computation will significantly increase when the dimension of the variable gets larger.
We present the training time and operational time in Table~\ref{tab:Table 10} for illustration. As shown, the training time of FCS is larger than that of QR models, but manageable compared to that of DeepAR.
Therefore, it is required to find computationally efficient methods to implement the proposed approach. As the fully conditional specification method iteratively estimates several conditional distribution models, it is hard to further reduce the training time. But it is feasible to directly learn the joint probability distribution model via a joint modeling approach, which would considerably reduce the training time.

\begin{table}[!ht]
\caption{Training time and operational time for 1-step probabilistic forecasting in Case~1.}
\small
 \label{tab:Table 10}
 \centering
 \begin{tabular}{cccccc}
 \hline
 & Gauss-R  & QR-R & Copula & FCS & DeepAR \\ 
 \hline
 Training time (min)  & 32 & 1 & 9 & 41 & 67 \\ 
 Operational time (s) & $\ll 0.01$ & $\ll 0.01$ & $\ll 0.01$ & 0.01 & 0.01 \\
 \hline
 \end{tabular}
\end{table}

\section{Conclusions}
It is intuitive to want to consider an ``impute, then predict" approach to deal with missing values, as existing forecasting methods can be readily used after the (imputing) pre-processing procedure. However, while such a pre-processing procedure at the model estimation stage jointly imputes input features and targets, it only imputes input features at the operational forecasting stage, possibly in a way that is not consistent with the model used for forecasting eventually. In this paper instead, we propose a ``universal imputation" approach, motivated by the problem of wind power forecasting in the presence of missing values. As for many other application areas, it is very common to have missing values within wind power forecasting. Our proposal approach relies on multiple imputation methods, and jointly performs the imputation of missing values of input features and the forecasting of targets. That is, it does not require a pre-processing procedure, while being consistent through model estimation and operational forecasting stages. Under the assumption that observations are missing at random, parameters can be estimated based on observations only, at the model estimation stage. At the operational stage, it treats targets as missing values and iteratively imputes both the missing values of input features and targets. Particularly, as multiple imputation provides several realizations from the joint distribution of input features and targets, the proposed approach naturally allows issuing both point and probabilistic forecasts. The case studies based on WIND Toolkit (over the USA) confirm the applicability of this approach. Not surprisingly, forecast quality necessarily decreases as the missing rate of the dataset increases. The results also suggest that the FCS-based method performs better than the ``impute, then predict" approach; it is especially preferred in the probabilistic forecasting case. And, the results suggest that the FCS-based approach may prevent overfitting to some extent. It also further validates the benefits from sharing information and data among wind farms, even in the presence of missing values.

We note that the modeling approach is quite different from the commonly used forecasting approaches in the context of complete datasets. The goal of this paper is not to replace the existing approaches, but to offer a complementary tool for use in the presence of missing values. We also expect there are similar ways to generalize commonly used modeling and forecasting approaches to the case of missing data. The computational costs of the introduced FCS-based approach are high and grow significantly as the dimension increases. Therefore, more efficient methods are still needed. It may be appealing to alternate the FCS method with distribution-free joint modeling imputation approaches. Our proposal is based on the ``missing-at-random'' assumption and thus avoids modeling the distribution of missingness. The situation where observations are missing not at random should be further explored in the future. Besides, emphasis should be placed on relaxing the stationary assumption in order to deal with non-stationary environments, e.g., with online learning.

\section*{Acknowledgments}

This work was performed during a research stay at the Technical University of Denmark. The authors would like to appreciate China Scholarship Council (NO. 202006230261). The research leading to this work is being carried out as a part of the Smart4RES project (European Union’s Horizon 2020, No. 864337). The sole responsibility of this publication lies with the authors. The European Union is not responsible for any use that may be made of the information contained therein. Besides, the authors would like to appreciate the reviewers and editors for their constructive suggestions.

%% If you have bibdatabase file and want bibtex to generate the
%% bibitems, please use
%%
 %\bibliographystyle{elsarticle-num} 
 \bibliographystyle{elsarticle-harv} 
 \bibliography{cas-refs,mylib}

%% else use the following coding to input the bibitems directly in the
%% TeX file.

% \begin{thebibliography}{00}

% %% \bibitem{label}
% %% Text of bibliographic item

% \bibitem{}

% \end{thebibliography}
\end{document}